%% file: SCAPA.tex
\documentclass[12pt,twoside]{article}
\title{Real Time Anomaly Detection And Categorisation}

\author{%
    Alexander T. M. Fisch \footremember{stor}{STOR-i Centre for Doctoral Training, Lancaster University, Lancaster LA1 4YF, U.K.} 
    \and Lawrence Bardwell \footremember{lanc1}{Department of Mathematics and Statistics, Lancaster University, Lancaster LA1 4YF, U.K.}%
    \and Idris A. Eckley \footrecall{lanc1}
  }
  
\date{\today}  
\input{_TeXdefs}

\graphicspath{{Figures/}}

\begin{document}
\bibliographystyle{apalike}  
\maketitle

\begin{abstract}
  The ability to quickly and accurately detect anomalous structure within data sequences is an inference challenge of growing importance. This work extends recently proposed post-hoc (offline) anomaly detection methodology to the sequential setting. The resultant procedure is capable of real-time analysis and categorisation between baseline and two forms of anomalous structure: point and collective anomalies. Various theoretical properties of the procedure are derived. These, together with an extensive simulation study, highlight that the average run length to false alarm and the average detection delay of the proposed online algorithm are very close to that of the offline version. Experiments on simulated and real data are provided to demonstrate the benefits of the proposed method.
\end{abstract}

\input{./sections/intro_Alex}
\input{./sections/background}

\input{./sections/CAPAonline_AlexV2}
\input{./sections/theory}
\input{./sections/sim_study_Alex}

\input{./sections/machine_temp}
\input{./sections/Acknowledgements}
\bibliography{bibliog}
\clearpage
\input{./sections/Supplementary.tex}

\end{document}

%% file: _TeXdefs.tex
\usepackage[left = 2.5cm, right = 2.5cm, top = 2cm, bottom = 2.5cm]{geometry} 
\usepackage{setspace}
\usepackage{graphicx}
\usepackage{amsfonts} 
\usepackage{amsmath}
\usepackage{float}
\restylefloat{table}
\usepackage{natbib} 
\usepackage{placeins}
\usepackage{caption}
\usepackage{subcaption}
\usepackage{caption}
\usepackage{amsthm}
\numberwithin{equation}{section}
\usepackage{algorithm}
\usepackage{algpseudocode}
\usepackage{titlepic}
\DeclareMathAlphabet{\mathpzc}{OT1}{pzc}{m}{it}

\usepackage{color}
\usepackage[toc,page]{appendix}
\usepackage{fancyhdr} 
\usepackage{url}
\usepackage{xcolor}
\usepackage{bbm}
\usepackage{verbatim}
\usepackage{multirow}
\usepackage{scrextend}

\newcommand{\footremember}[2]{%
   \footnote{#2}
    \newcounter{#1}
    \setcounter{#1}{\value{footnote}}%
}
\newcommand{\footrecall}[1]{%
    \footnotemark[\value{#1}]%
}

\algnewcommand{\Inputs}[1]{%
 \State \textbf{Inputs:}
 \Statex \hspace*{\algorithmicindent}\parbox[t]{.8\linewidth}{\raggedright #1}
}
\algnewcommand{\Initialize}[1]{%
 \State \textbf{Initialize:}
 \Statex \hspace*{\algorithmicindent}\parbox[t]{.8\linewidth}{\raggedright #1}
}
\makeatletter
\algrenewcommand\ALG@beginalgorithmic{\footnotesize}
\makeatother

\newtheorem{Thm}{Proposition}

\DeclareMathOperator*{\argmin}{argmin}

\algnewcommand\algorithmicswitch{\textbf{switch}}
\algnewcommand\algorithmiccase{\textbf{case}}
\algnewcommand\algorithmicassert{\texttt{assert}}
\algnewcommand\Assert[1]{\State \algorithmicassert(#1)}%
\algdef{SE}[SWITCH]{Switch}{EndSwitch}[1]{\algorithmicswitch\ #1\ \algorithmicdo}{\algorithmicend\ \algorithmicswitch}%
\algdef{SE}[CASE]{Case}{EndCase}[1]{\algorithmiccase\ #1}{\algorithmicend\ \algorithmiccase}%
\algtext*{EndSwitch}%
\algtext*{EndCase}%

\usepackage[pdftex]{hyperref}  

%% file: sections/intro_Alex.tex
\section{Introduction}
\label{sec:intro}

The detection of anomalies in time series has received considerable attention in both the statistics \citep{doi:10.1080/01621459.1993.10594321} and machine learning \citep{Chandola:2009:ADS:1541880.1541882} literature. This is no surprise given the broad range of applications from fraud detection \citep{Ferdousi2006UnsupervisedOD} to fault detection \citep{theissler2017detecting,zhao2018anomaly} that this area lends itself to. In recent years, the proliferation of sensors within the internet of things (IoT) has led to the emergence of real time detection of anomalies in streaming (high frequency) data as an important new challenge. 

\begin{figure}[h!]
  \begin{subfigure}[t]{0.33\linewidth}
    \includegraphics[scale=.33]{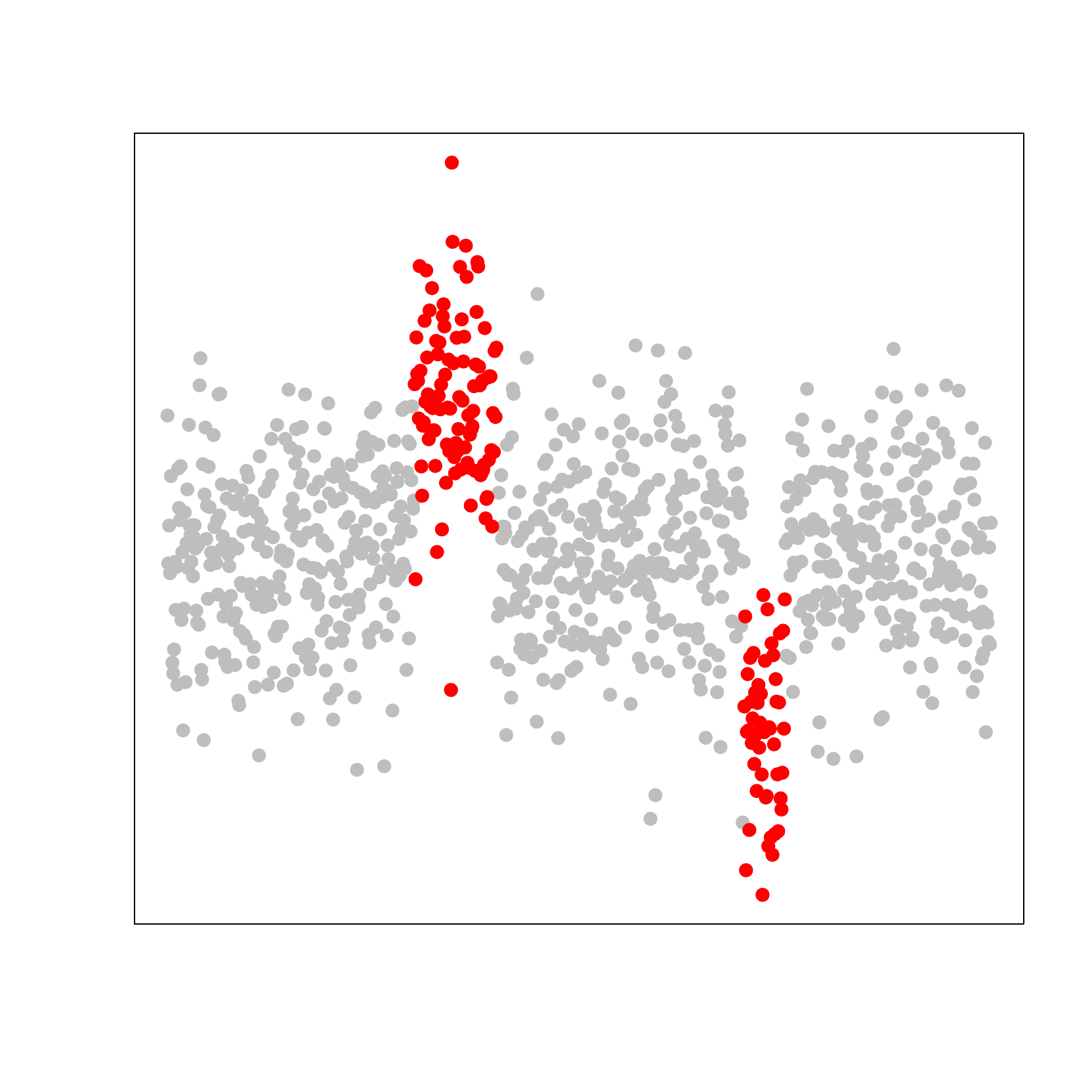}
  \end{subfigure}\hfill
  \begin{subfigure}[t]{0.33\linewidth}
    \includegraphics[scale=.33]{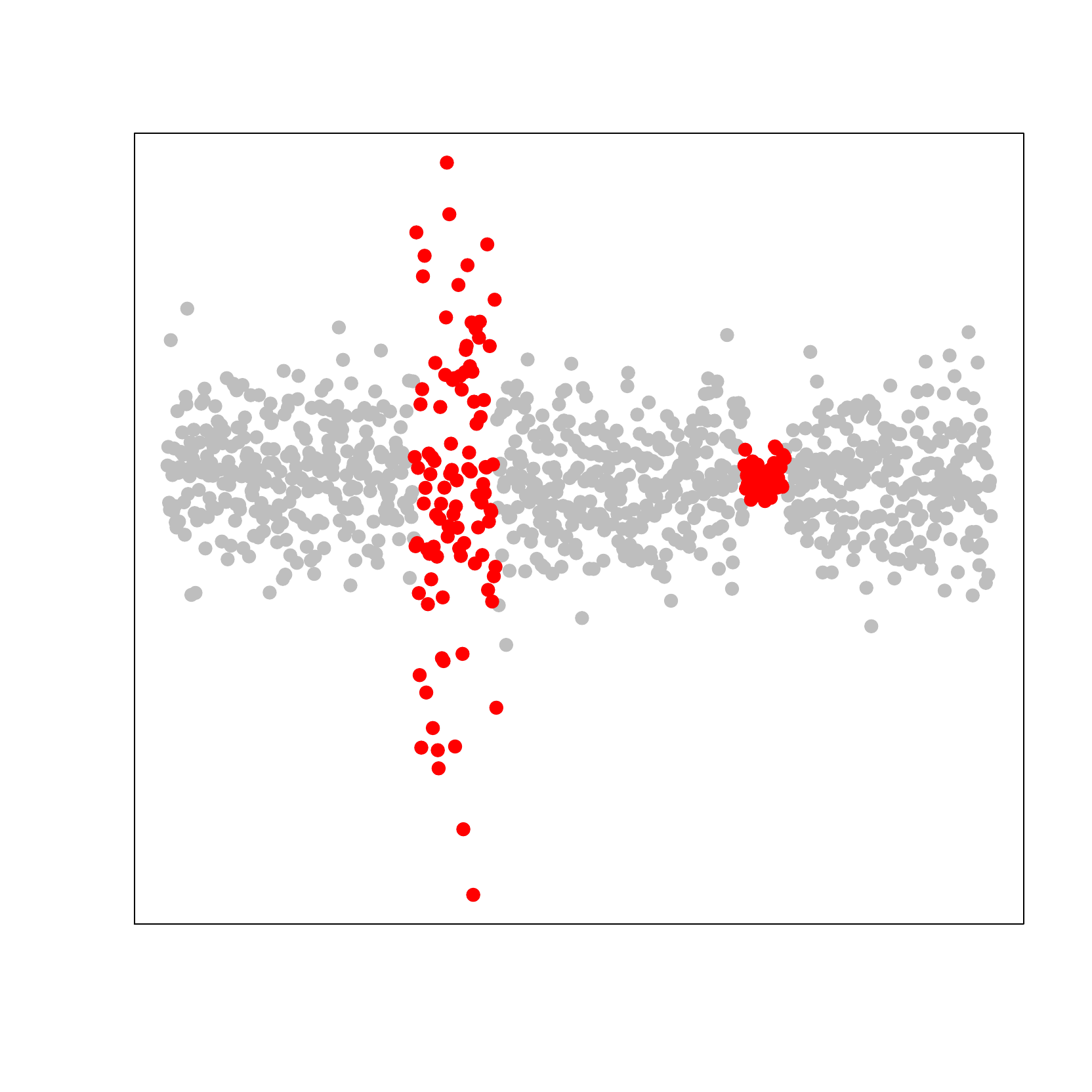}
  \end{subfigure}
  \hfill
  \begin{subfigure}[t]{0.33\linewidth}
    \includegraphics[scale=.33]{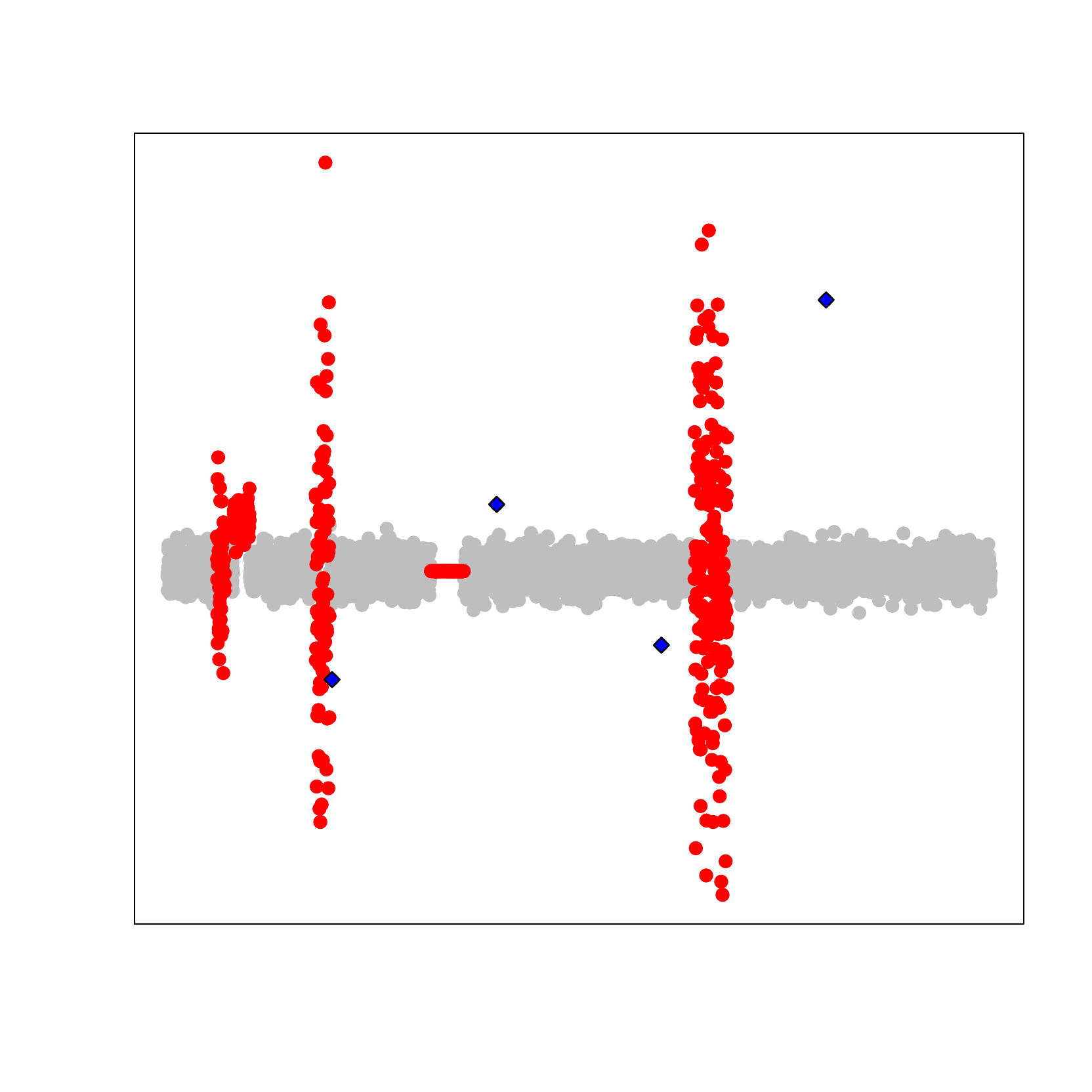} 
  \end{subfigure} 
\caption{Time series containing collective and point anomalies. Typical data shown in grey, anomalous segments in red and point anomalies shown in blue.}
\label{fig:main}
\end{figure}

Anomalies can be classified in a number of different ways \citep{Chandola:2009:ADS:1541880.1541882}. In this work, following the definitions of \citet{2018arXiv180601947F}, we distinguish between point and collective anomalies. Point anomalies, also known as outliers, global anomalies or contextual anomalies \citep{Chandola:2009:ADS:1541880.1541882}, are single observations that are anomalous with regards to their local or global data context. Conversely, collective anomalies, also known as abnormal regions \citep{bardwell2017}, or epidemic changepoints \citep{doi:10.1002/sim.4780040408}, are sequences of contiguous observations which are not necessarily anomalous when compared to either their local or the global data context but together form an anomalous pattern. Figure \ref{fig:main} provides several examples. In this paper, collective anomalies and epidemic changepoints are used interchangeably.


%
The epidemic changepoint model assumes that data follows some baseline, or typical distribution, everywhere except for some anomalous time windows during which it follows another distribution. The detection of epidemic changes in mean was first studied by \citet{doi:10.1002/sim.4780040408} with applications to epidemiology. Since then, research in this area has been driven by various applications including the detection of copy number variants in DNA \citep{doi:10.1093/biostatistics/kxh008,doi:10.1093/biomet/ass059,bardwell2017} and the analysis of brain imaging data \citep{aston2012,2019arXiv190300288S}. In particular, much of the pertinent literature has concentrated 
on the epidemic change in mean setting. See \citet{10.2307/2336767} and \citet{Gut2005} for details. 

More recently, the detection of joint epidemic changes in mean and variance as well as point anomalies was considered by \cite{2018arXiv180601947F}. 
%
In parallel, there has also been some work on detecting anomalies within the online setting. \citet{Gut2005} consider the problem of detecting epidemic changepoints sequentially while \cite{5990537} and \cite{AHMAD2017134} propose methods for the online detection of point anomalies. 

The main contribution of this paper is to extend the offline Collective And Point Anomaly (CAPA) algorithm of \cite{2018arXiv180601947F} to the online setting to detect both collective and point anomalies in streaming data, formalising early heuristic ideas appearing in \cite{BardwellConfPaper}.
We call this algorithm Sequential-CAPA (SCAPA).  To the best of our knowledge, SCAPA is the first statistical approach to jointly detect point and collective anomalies in an online fashion within the epidemic changepoint framework for unknown mean and variance.

The article is organised as follows. In Section \ref{sec:background} we introduce the literature on offline detection of anomalous time series regions, particularly focusing on the recently proposed CAPA approach. Section \ref{sec:online} proceeds to extend this methodology to the online setting, introducing the Sequential Collective and Point Anomaly (SCAPA) algorithm. Theoretical properties of the proposed methodology are investigated in Section \ref{sec:theory}. Further results, together with a set of simulation studies is given in Section \ref{sec:sims}, indicating how these can be used to inform practitioners on how to select the hyper-parameters of SCAPA. Finally, we apply SCAPA to the monitoring of a sensor on a publically available, industrial machine-level data in Section \ref{sec:temp}. All proofs can be found in the supplementary material.

%% file: sections/background.tex

\section{Background}
\label{sec:background}

CAPA, introduced by \citet{2018arXiv180601947F}, seeks to jointly detect and distinguish between point and collective anomalies within an offline, univariate time series setting. The heart of the approach is founded upon an epidemic changepoint model. To this end, consider a stochastic process $x_t \sim \mathcal{D}(\theta(t))$, drawn from some distribution, $\mathcal{D}$, indexed by a set of model parameters, $\theta(t)$. Collective anomalies can then be modelled as epidemic changes of the set of parameters $\theta(t)$. I.e.\ time windows in which $\theta(t)$ deviates from the typical, and potentially unknown, set of parameters $\theta_0$. Formally,
\begin{align*}
 \theta(t) = 
  \begin{cases}
    \theta_1  & s_1  < t \leq e_1 \\
    &\vdots \\
    \theta_K   & s_K < t \leq e_K \\
    \theta_0 & \text{otherwise.}
     \end{cases}
\end{align*}
Here $K$ denotes the number of collective anomalies, while $s_i$, $e_i$, and $\theta_i$ correspond to the start point, end point and the unknown parameter(s) of the $i$th collective anomaly respectively. 

The number and locations of collective anomalies are estimated by choosing $K, (s_1,e_1) , \hdots , (s_k,e_K)$, and $\theta_0$ such that they minimise the penalised cost 
\begin{align}
  \label{eq:costmin} \sum_{t \notin \cup [s_i+1,e_i] } \mathcal{C}(x_t,\theta_0) + \sum_{j=1}^{K} \left[ \min_{\theta_j} \left( \sum_{t=s_j +1}^{e_j} \mathcal{C}(x_t,\theta_j)  \right) + \beta_C \right].
\end{align}
$\mathcal{C}(\cdot,\cdot)$ is a cost function, e.g.\ twice the negative log-likelihood, and $\beta_C$ is a penalty term for introducing a collective anomaly, which seeks to prevent overfitting. A minimum segment length, $l$, can be imposed by adding the constraint $e_k - s_k \geq l$ for $k = 1,2, \hdots ,K$, if collective anomalies of interest are assumed to be of length at least $l \geq 1$.

Minimising the cost function \eqref{eq:costmin} exactly by solving a dynamic programme like the PELT method \citep{pelt} is not possible. This is because the parameter of the typical distribution, $\theta_0$, is shared across segments, and introduces dependence. \citet{2018arXiv180601947F} suggest removing this dependence in $\theta_0$ by obtaining a robust estimate $\hat{\theta}_0$ over the whole data and then minimising
\begin{align}
\label{eq:costminbis} \sum_{t \notin \cup [s_i+1,e_i] } \mathcal{C}(x_t,\hat{\theta}_0) + \sum_{j=1}^{K} \left[ \min_{\theta_j} \left( \sum_{t=s_j +1}^{e_j} \mathcal{C}(x_t,\theta_j)  \right) + \beta_C \right],
\end{align}
as an approximation to \eqref{eq:costmin} over just the number and location of collective anomalies. The main focus of \citet{2018arXiv180601947F} was on the case where anomalies are characterised by an atypical mean and or variance. In this case, the authors suggest minimising \small
\begin{align*}
\sum_{t \notin \cup [s_i+1,e_i] } \left[ \log (\hat{\sigma}_0^2) + \left( \frac{x_t-\hat{\mu}_0}{\hat{\sigma}_0} \right)^2 \right] +
\sum_{j=1}^{K} \left[ (e_j - s_j)\left( \log \left( \frac{ \sum_{t=s_j + 1}^{e_j} (x_t - \bar{x}_{(s_j + 1):e_j})^2}{(e_j-s_j)} \right) + 1 \right) + \beta_{C}  \right],
\end{align*}
\normalsize
subject to a minimum segment length $l$ of at least 2. The above expression arises from setting the cost function to twice the negative log-likelihood of the Gaussian. The robust estimates for mean and variance, $\hat{\mu}_0$ and $\hat{\sigma}_0$, can be obtained from the median and the inter-quartile range.
 
The main weakness of the above penalised cost is that point anomalies will be fitted as collective anomalies in a segment of length $l$. To remedy this, point anomalies are modelled as epidemic changes of length one in variance (only). The set of point anomalies is denoted as $O$. To infer both collective and point anomalies we minimise
\begin{align}
  \label{eq:CAPA}
  \begin{split}\sum_{t \notin \cup [s_i+1,e_i] \cup O } \left[ \log (\hat{\sigma}_0^2) + \left( \frac{x_t-\hat{\mu}_0}{\hat{\sigma}_0} \right)^2 \right] + \sum_{t \in O} \left[ \log( (x_t - \hat{\mu}_0)^2 ) + 1 + \beta_O \right] + \\
  \sum_{j=1}^{K} \left[ (e_j - s_j)\left( \log \left( \frac{ \sum_{t=s_j + 1}^{e_j} (x_t - \bar{x}_{(s_j + 1):e_j})^2}{(e_j-s_j)} \right) + 1 \right) + \beta_{C}  \right],
  \end{split}
 \end{align}
 with respect to $K, (s_1,e_1) , \hdots , (s_k,e_K)$, and $O$, subject to the constraint $e_k - s_k \geq l \geq 2$ for $k = 1,2, \hdots , K$. Here, $\beta_O$ corresponds to a penalty for a point anomaly. 

The CAPA algorithm then minimises the cost in \eqref{eq:CAPA} by solving the dynamic programme 
\begin{align*}
C(t) = \min \Bigg[ &C(t-1) + \left( \frac{x_t-\hat{\mu}_0}{\hat{\sigma}_0} \right)^2  , C(t-1) + \log\left( \left(x_t - \hat{\mu}_0\right)^2 \right) + 1 + \beta_O, \\
&\min_{0 \leq k < t-l} \left( C(k) + (t - k)\left( \log \left( \frac{ \sum_{i=k+1}^{t} (x_i - \bar{x}_{(k+1):t})^2}{(t-k)} \right) + 1 \right) \right) + \beta_C  \Bigg],
\end{align*}
taking $C(0) = 0$.

%% file: sections/CAPAonline_AlexV2.tex
\section{Sequential CAPA}
\label{sec:online}

We now introduce our Sequential CAPA procedure. In extending CAPA to the online setting three main challenges arise. Specifically, any approach developed should be mindful of the following: (i) that the computational and storage cost of the dynamic programme increase with time; (ii) the typical (baseline) parameters have to be learned online and (iii) penalty selection.
We address each of these three challenges in turn, proposing solutions in the following sections, prior to formally introducing the SCAPA algorithm in Section \ref{sec:SCAPA}.

\subsection{Increasing Computational And Storage Cost}

As noted in Section \ref{sec:background}, CAPA infers collective and point anomalies by solving a set of dynamic programme recursions. However both the computational cost of each recursion, and the storage cost, increase linearly in the total number of observations. This is unsuitable for the online setting in which both storage and computational resources are finite. 

In practice, this problem can be surmounted by imposing a maximum length $m$ for collective anomalies. This can be achieved by adding the set of constraints
\begin{align}
  \label{eq:constraint}
  e_i - s_i \leq m \hspace{10pt} \forall i = 1,2, \hdots ,K 
\end{align}
to the optimisation problem in equation \eqref{eq:CAPA}. The resulting problem can then be solved using the following dynamic programme
\begin{align*}
C(t) = \min \Bigg[ &C(t-1) + \left( \frac{x_t-\hat{\mu}_0}{\hat{\sigma}_0} \right)^2  , C(t-1) + \log\left( \left(x_t - \hat{\mu}_0\right)^2 \right) + 1 + \beta_O, \\
&\min_{t-m \leq k < t-l} \left( C(k) + (t - k)\left( \log \left( \frac{ \sum_{i=k+1}^{t} (x_i - \bar{x}_{(k+1):t})^2}{(t-k)}\right) + 1 \right) \right) + \beta_C  \Bigg].
\end{align*}

As a consequence of restriction \eqref{eq:constraint}, each recursion only requires a finite number of calculations. Moreover, only a finite number of the optimal costs, $C(t)$, need to be stored in memory. The practical implications of this additional constraint are likely to be limited. Within this setting collective anomalies encompassing fewer than $m$ observations will be detected as before. However, for those scenarios where an anomaly encompasses more than $m$ observations, these will be fitted as a succession of collective anomalies each of length less than $m$, provided that their signal strength (cf Section \ref{sec:sims} for a definition) is large enough. As one might anticipate, within this setting long anomalous segments with low signal strength would not be detectable any more as a result of the approximation. 

\subsection{Sequential Estimation Of The Typical Parameters}

As described in Section \ref{sec:background}, the dynamic programme used by CAPA requires robust estimates of the set of typical parameters $\theta_0 = (\mu_0,\sigma_0)$. \citet{2018arXiv180601947F} estimate $\mu_0$ and $\sigma_0$ on the full data using the median and inter-quartile range respectively. In an online setting, however, these quantiles have to be learnt as the data is observed.

A range of methods have been proposed that aim to  
estimate the cumulative distribution function (CDF) of the data sequentially and use it to estimate quantiles. For example, \citet{doi:10.1137/0904048} proposed a method based on techniques from Stochastic Approximation (SA) to estimate the $\alpha$th quantile $x_{(\alpha)}$ of an unknown distribution function. Moreover, \citet{doi:10.1137/0904048} also established that, in the i.i.d.\ setting, the resulting sequential estimates $\hat{x}_{(\alpha),n} \rightarrow x_{(\alpha)}$ almost surely as the number of observations $n \rightarrow \infty$. Under the same assumptions, they also showed that $\sqrt{n}( \hat{x}_{(\alpha),n} - x_{(\alpha)})$ converges in distribution to a Normal distribution. These consistency results are important for an online implementation of CAPA, as \cite{2018arXiv180601947F} showed that the consistency of CAPA requires the robustly estimated mean and variance to be within $O_p\left(\sqrt{\frac{\log(n)}{n}}\right)$ of the true typical mean and variance.

\begin{figure}[!h] 
	\begin{subfigure}[b]{\linewidth}
		\centering
		\includegraphics[width=0.65\linewidth]{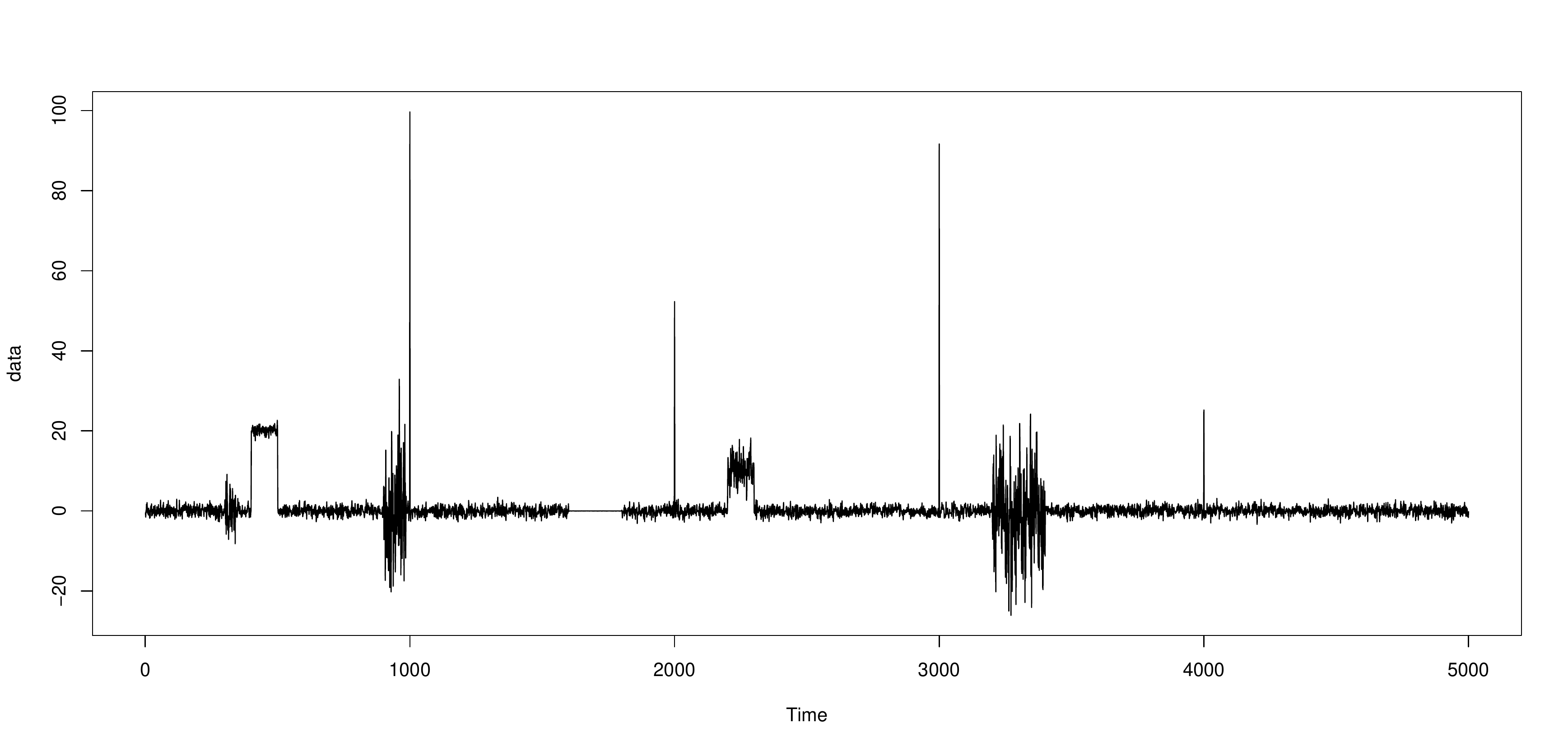} 
		\caption{Example time series} 
	\end{subfigure} 
    \begin{subfigure}[b]{\linewidth}
		\centering
		\includegraphics[width=0.65\linewidth]{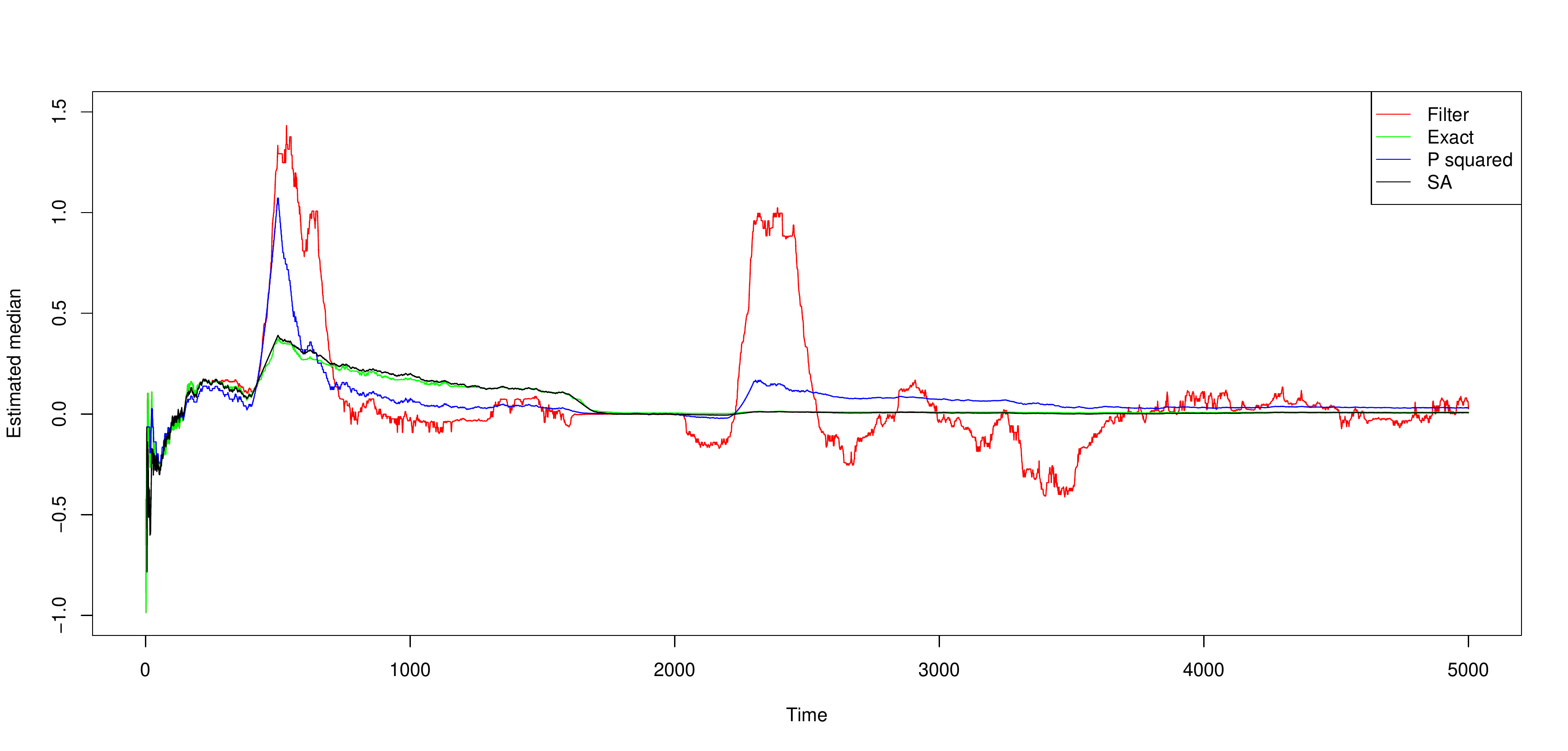} 
		\caption{Sequentially estimated median} 
	\end{subfigure} 
		\begin{subfigure}[b]{\linewidth}
		\centering
		\includegraphics[width=0.65\linewidth]{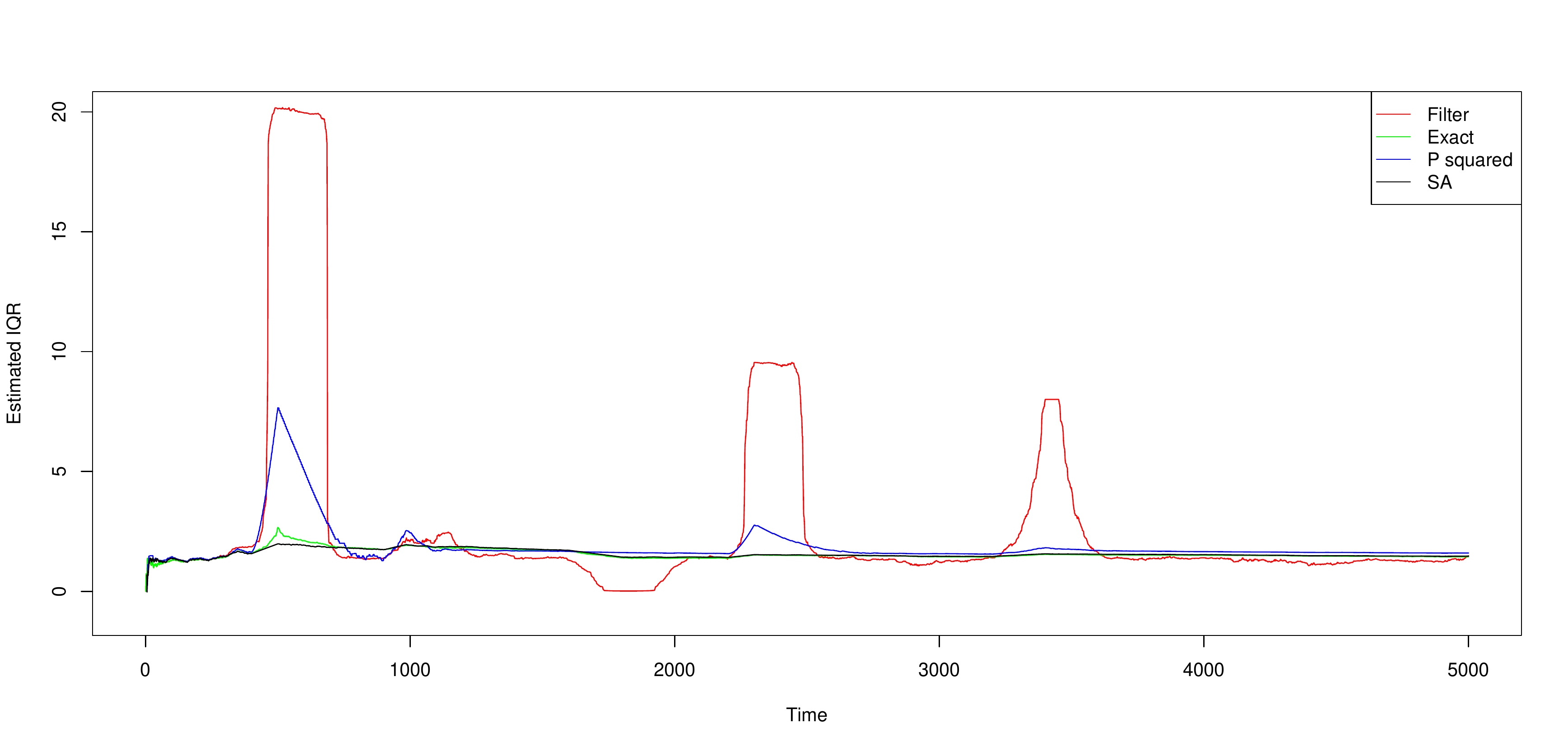} 
		\caption{Sequentially estimated IQR} 
	\end{subfigure} 
	  \caption{a) Example time series with collective and point anomalies as well as the b) median and c) IQR estimated sequentially over time using different methods: The quantile filter by \citet{Justusson1981} (Filter), the $p^2$-method of \citet{Jain:1985:PAD:4372.4378} (P squared) and the Stochastic Approximation based method by  \citet{doi:10.1137/0904048} (SA).}
  \label{fig:example_series}
\end{figure}

The memory required to obtain the SA-estimate is finite and small. Moreover, the standard errors of the SA-estimate and sample quantiles are close even for relatively small sample sizes, as can be seen from Figure \ref{fig:example_series}. Further, we note that these estimates tend to be considerably more accurate than those of other commonly used methods such as the quantile filter \citep{Justusson1981} and the $p^2$-algorithm \citep{Jain:1985:PAD:4372.4378}. This is due to the fact that the quantile filter is not consistent, and that the $p^2$-algorithm  is not robust with respect to outliers, thus losing a critical property of quantile estimators. 

Pseudo-code for the SA-based method is given in Algorithm \ref{alg:SQE}. Using a burn in period to stabilise the quantile estimates is recommended, as even the exact order statistics take some time to initially converge. SA-based methods can also be used to calculate other important statistics in an online fashion. For example, \citet{Sharia2010} applied SA-techniques to learn auto-regressive parameters sequentially. Such estimators can be used to inflate the penalties used to account for deviations from the i.i.d.\ assumptions. This is discussed in more detail in Section \ref{sec:sims}.

\subsection{Penalty Selection}
We now turn to the important question of penalty selection. In the offline setting, penalties are typically chosen to control false positives under the null hypothesis. For example, \citet{2018arXiv180601947F} suggested using penalties 
\begin{align}
	\label{eq:pens_t}
		\beta_C(a,\lambda)  = 2\frac{a}{a-1} \left( 1 + \lambda + \sqrt{2\lambda} \right), \;\;\;\;\;\;\;\;\;\;\;\;\;\;\;\;\;\;\;\;\;\;\;\;
		 \beta_O(\lambda)  = 2\lambda, \;\;\;\;\;\;\;\;\;\;\;\;\;\;
\end{align}
indexed by a single parameter $\lambda$ for CAPA when considering the change in mean and variance setting. Here, the penalty for collective anomalies, $\beta_C$, depends on the length $a$ of the putative collective anomaly. The motivation for these penalties is to ensure that the estimates for the number of collective anomalies and the set of point anomalies,  $\hat{K}$ and $\hat{O}$, satisfy
\begin{align}
	\Pr( \hat{K} = 0 , \hat{O} = \emptyset ) \geq 1 - C_1ne^{-\lambda} - C_2(ne^{-\lambda})^2,
\end{align}
under the null hypothesis that no point or collective anomaly is present in the data. Consequently, setting $\lambda = \log(n)$ asymptotically controls the number of false positives of a time series of length $n$.

In the online setting, however, the concept of the length of a time series does not exist. Consequently, fixed constants are used for the penalties instead. This means that, unless the errors are bounded, false positives will be observed eventually. In common with \citet{lorden1971} and \citet{pollak1985}, we suggest choosing $\lambda$ to be as small as possible, to maximise power against anomalies, whilst maintaining the average run length (ARL), the average time between false positives, at an acceptable level. Practical guidance on the choice of $\lambda$ can be taken from Proposition \ref{Thm:ARL}, which provides an asymptotic result for the relationship between the log-ARL and $\lambda$, under a certain model form. This relationship is empirically verified for other models using simulations in Section \ref{sec:sims}.

\subsection{Sequential Collective And Point Anomaly} \label{sec:SCAPA}

Given the above solutions to the three identified challenges, we are able to extend CAPA to an online setting. We call the resultant approach Sequential Collective And Point Anomaly (SCAPA). The basic steps of the algorithm are as follows: When an observation comes in, it is used to update the sequential estimates of the typical parameters. The observation is then standardised using the typical mean and variance $(\mu_0, \sigma^2_0)$, before being passed to the finite horizon dynamic programme. Detailed pseudocode can be found in Algorithm \ref{alg:OCAPA} of the supplementary material

\begin{figure}
  \begin{subfigure}[t]{0.33\linewidth}
    \includegraphics[scale=.33]{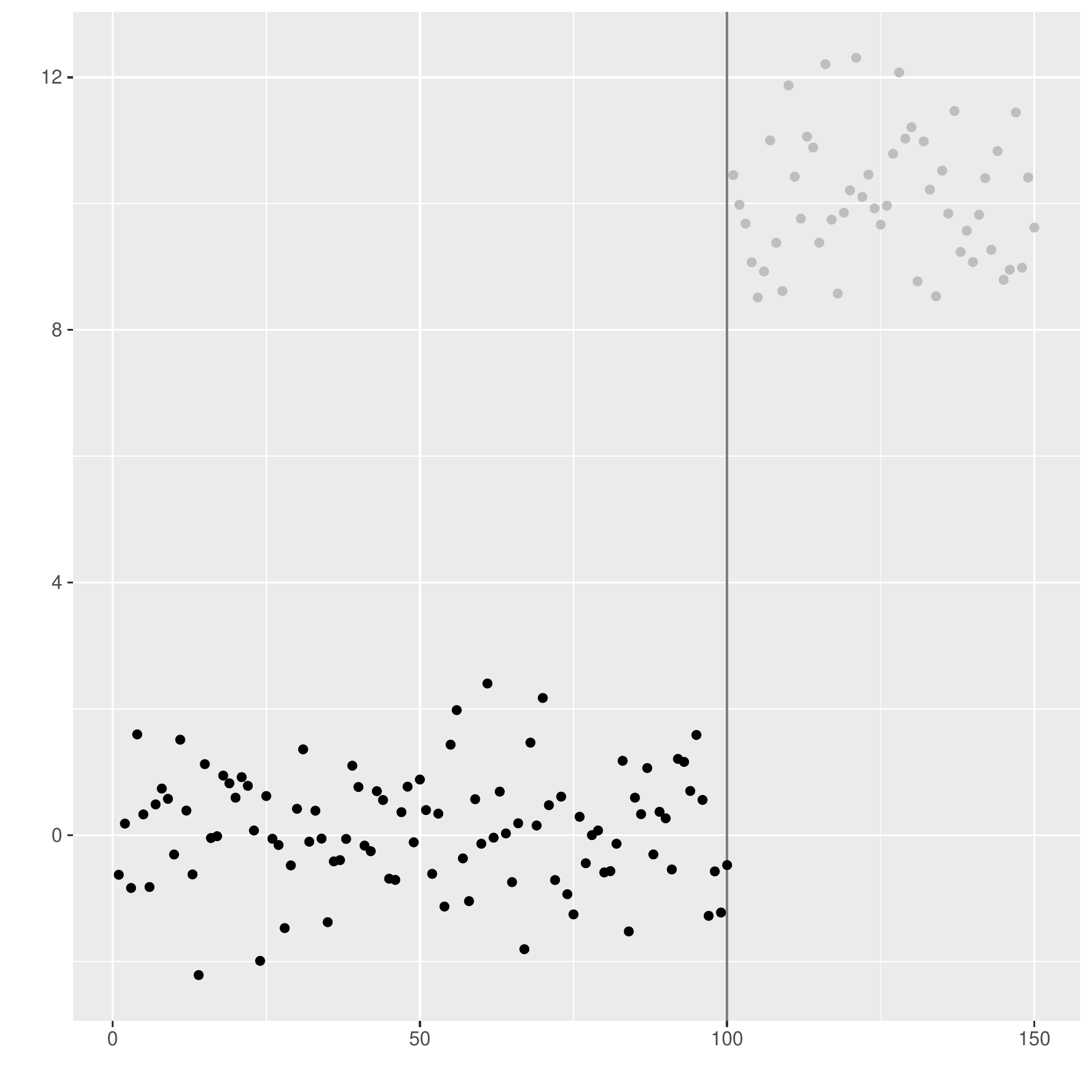}
    \caption{}
  \end{subfigure}\hfill
  \begin{subfigure}[t]{0.33\linewidth}
    \includegraphics[scale=.33]{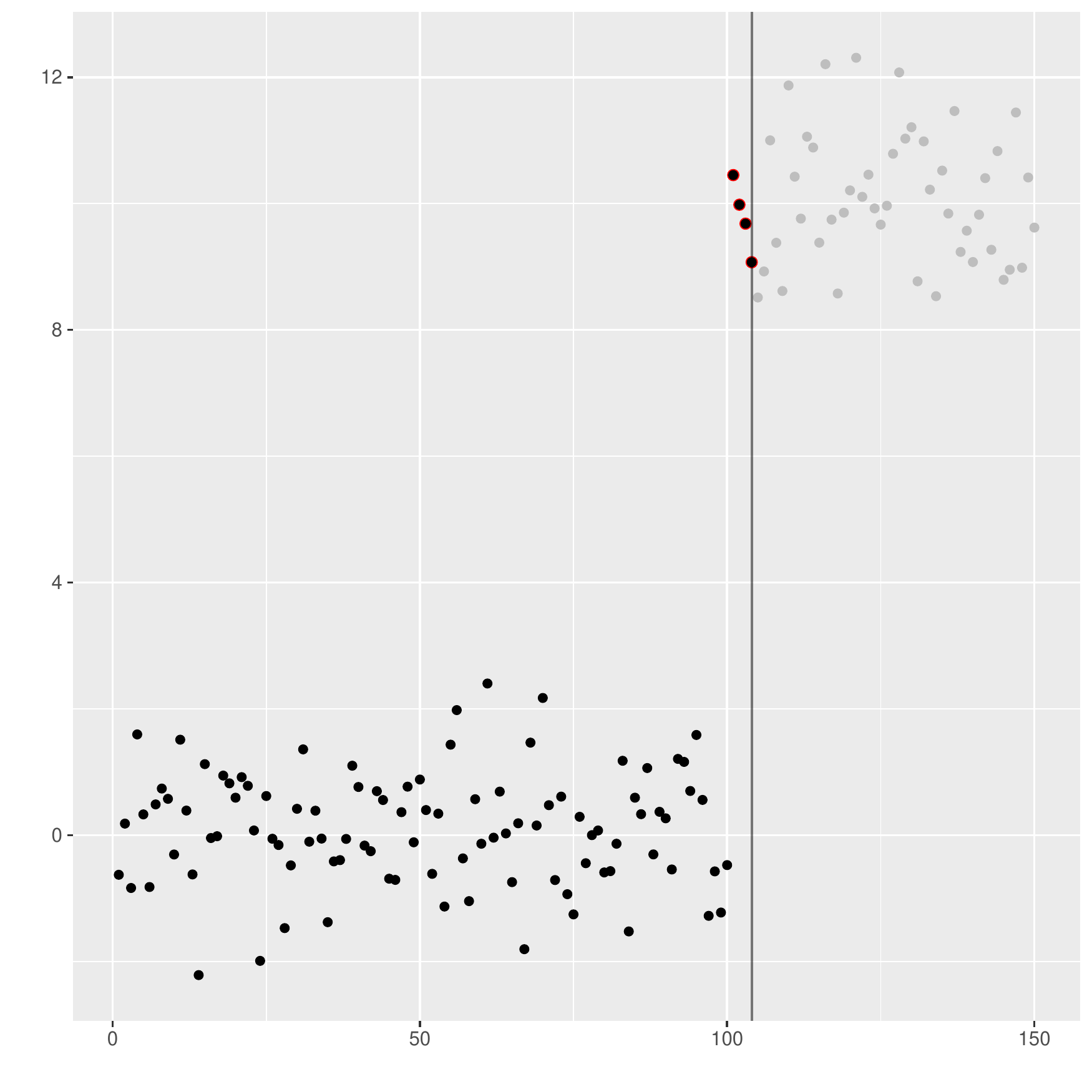}
    \caption{}
  \end{subfigure}
  \hfill
  \begin{subfigure}[t]{0.33\linewidth}
    \includegraphics[scale=.33]{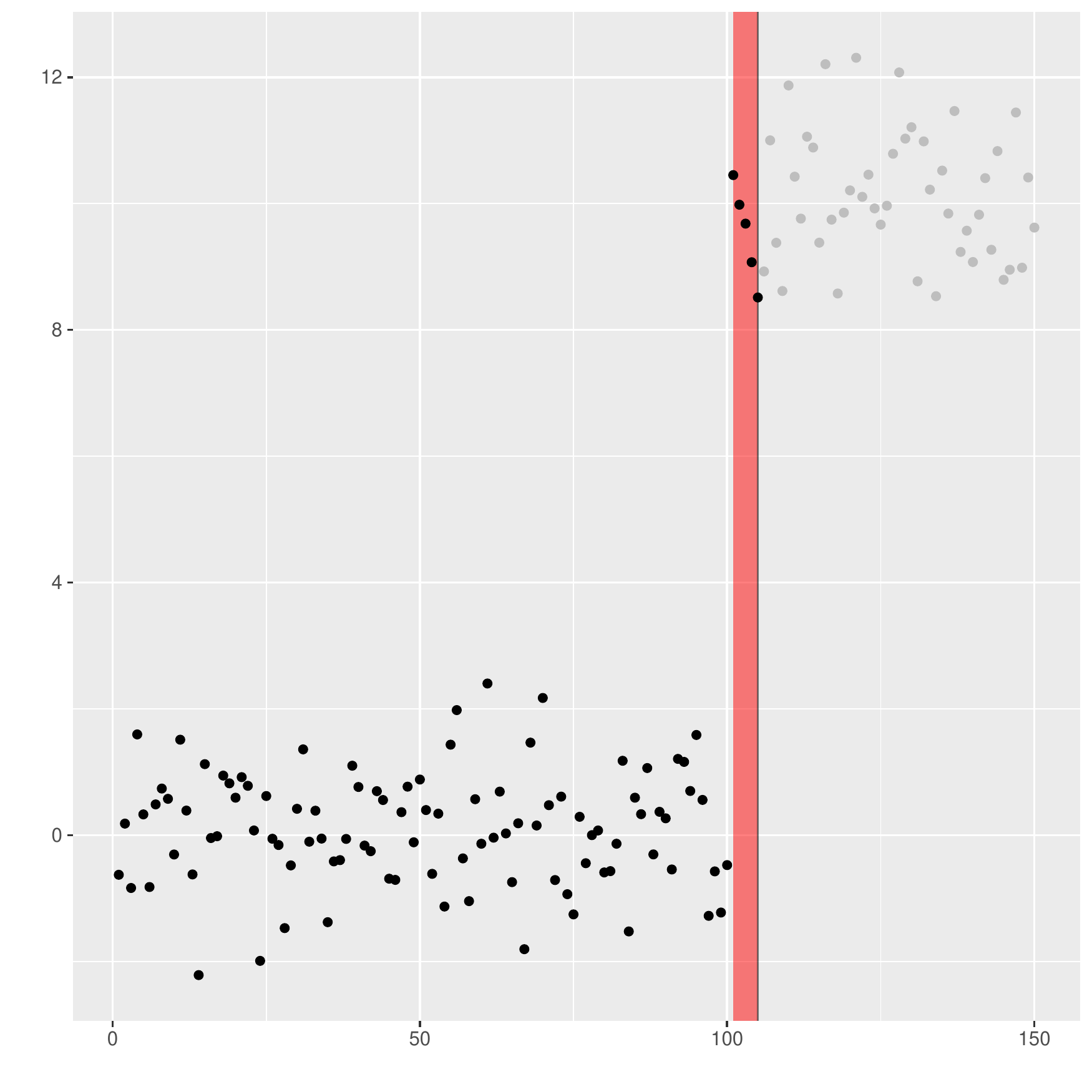}
    \caption{}
  \end{subfigure}  
\caption{The evolution in the detection of a collective anomaly with a minimum segment length of $l = 5$. The times shown are a) $t = 100$ just prior to the anomalous observations, b) $t = 104$ where the observations $x_{101:104}$ have been labelled as point anomalies and c) $t = 105$ where the observations $x_{101:105}$ have been labelled as a collective anomaly.}
\label{fig:SCAPA_sequential}
\end{figure}

The sequential nature of SCAPA's analysis is displayed in Figure \ref{fig:SCAPA_sequential} across three plots, each representing the output of the analysis at different time points. Note how a collective anomaly is detected, initially, as a sequence of point anomalies until the number of observations equals the minimum segment length.

%% file: sections/theory.tex
\section{Theory}
\label{sec:theory}

We now turn to consider the theoretical properties of SCAPA. In particular, we investigate the average run length (ARL) and the average detection delay (ADD). Here, the ARL corresponds to the expected number of baseline datapoints SCAPA processes before detecting a false positive. Conversely, the ADD corresponds to the expected number of observations between the onset of a collective anomaly and the time at which a collective anomaly is first detected. We will place a particular emphasis on the effects of the maximum segment length, $m$ on the ADD, as the results following from that analysis provide practical guidance on how to choose $m$. 

For simplicity of exposition, we will restrict our attention to the change in mean setting, in which the penalised cost is 
\begin{align*}
\sum_{t \notin \cup [s_i+1,e_i] \cup O }\left( \frac{x_t-{\mu}_0}{{\sigma}_0} \right)^2 + \sum_{t \in O} \left[ 0 + \beta_O \right] + 
\sum_{j=1}^{K} \left[  \sum_{t=s_j + 1}^{e_j} \left( \frac{  x_t - \bar{x}_{(s_j + 1):e_j}}{{\sigma}_0} \right)^2  + \beta_{C}  \right]
\end{align*}


In this setting, the ARL of SCAPA can be related to the penalty constant, $\lambda$, via the following result:
\begin{Thm}\label{Thm:ARL}
	Assume we observe a data sequence with typical mean, $\mu_0$, and the typical variance, $\sigma_0^2$, both known. Then the ARL of SCAPA on i.i.d.\ $N(\mu_0,\sigma_0^2)$-distributed observations $x_1,x_2,...$ then satisfies
	\begin{equation*}
	\log(ARL) \sim \lambda/2
	\end{equation*}
	as $\lambda \rightarrow \infty$.
\end{Thm}

\textbf{Proof:} See appendix.

As a consequence of the above, the probability of false alarm is proportional to $\exp(-\lambda/2)$. As discussed in the previous section, this can be used to inform the choice of penalty in practice if an acceptable probability of false alarm is given.


We now turn to investigate the effects of the maximum segment length, $m$, on the ADD. 
To simplify the exposition of these results, we assume that the collective anomaly begins at time $\tau = 0$. Formally, consider the series
\begin{equation}\label{eq:model}
x_1, x_2, ... \stackrel{i.i.d.}{\sim} N(\mu,1)
\end{equation}
and assume that the typical mean, $\mu_0$, is equal to $0$ and known. For a maximum segment length $m$, we then define $ADD_m$ to be the ADD of SCAPA with a maximum segment length $m$. Additionally, we define $ADD_\infty$ to be the ADD of SCAPA without maximum segment length. The following proposition shows that imposing a maximum segment length does not affect the ADD, provided that the maximum segment length increases at a rate faster than the penalty.
\begin{Thm}\label{Thm:ADD_max}
	Let $x_1,x_2,.. $ follow the distribution specified in \eqref{eq:model}. Moreover, let the known baseline mean and variance be 0 and 1 respectively. Then, if $m > \frac{\lambda}{\mu^2}(1+\epsilon)$ for some $\epsilon >0$,
	\begin{equation*}
	ADD_m = ADD_\infty + o(1)
	\end{equation*}
	as $\lambda \rightarrow \infty$.
\end{Thm}

\textbf{Proof:} See appendix.

Given Proposition \ref{Thm:ADD_max}, it is natural to consider what happens in the converse setting. I.e.\ what happens if the maximum segment length increases at a slower rate than the penalty. 

\begin{Thm}\label{Thm:ADD_min}
	Let $x_1,x_2,.. $ follow the distribution specified in \eqref{eq:model}. Moreover, let the known typical mean and variance be 0 and 1 respectively. Then, if $1 \leq m < \lambda^{1-\epsilon}$ for some $\epsilon >0$
	\begin{equation*}
	\log(ADD_m) \sim \lambda/2
	\end{equation*}
	as $\lambda \rightarrow \infty$.
\end{Thm}

\textbf{Proof:} See appendix.

In other words, the log-ADD has the same exponential rate as the log-ARL on non-anomalous data.

As previously discussed, limits on the number of possible interventions often determine a tolerable probability of false alarm in practice. Proposition \ref{Thm:ARL} therefore provides a mechanism to determine a suitable penalty constant $\lambda$. Further, Propositions \ref{Thm:ADD_max} and \ref{Thm:ADD_min} can be used to help inform an appropriate choice of maximum segment length, $m$. Specifically $m$ should be at least of magnitude $\frac{\lambda}{\mu^2}$, where $\mu$ is the smallest change in mean of interest to ensure power.


%% file: sections/sim_study_Alex.tex
\clearpage
\newpage
\section{Simulation Study}
\label{sec:sims}

We now turn to examine the performance of SCAPA in various simulated settings. 
We start by considering the case where a single collective anomaly is present to evaluate SCAPA via its ARL and ADD performance in Section \ref{sec:single}. The effect of auto-correlation is also examined. This is followed by a comparison with CAPA on time series containing multiple anomalies in Section \ref{sec:Multi}.

\subsection{A Single Anomaly} \label{sec:single}

Prior to describing our first simulation scenario, we begin by noting that the ARL and ADD are functions of $\beta_C$ and $\beta_O$. Further, as we have seen in equation \eqref{eq:pens_t} these are a function of a single parameter $\lambda$. The aim of our simulation study, therefore, is to inform the choice of $\lambda$ that gives a suitable ARL/ADD trade off. In particular, \textit{ceteri paribus}, a weaker change gives rise to a larger delay than a stronger change. In other words, we must control for the strength of change when investigating the ADD. To do so, we take the definition of signal strength from \citet{2018arXiv180601947F}. 

For a collective anomaly with mean $\mu$ and variance $\sigma^2$ the strength, $\Delta$, of a change is defined as 
\begin{align*}
  \Delta = \log \left( 1 + \frac{1}{2}\Delta_\sigma^2 + \frac{1}{4}\Delta_\mu^2 \right)  \;\;\;\;\;\;
  \;\;\;\;\;\; \Delta_\mu^2 = \frac{  (\mu_0 - \mu)^2 }{\sigma_0\sigma}   \;\;\;\;\;\;\;\;\;\;\;\;
  \Delta_\sigma^2 = \frac{\sigma_0}{\sigma} + \frac{\sigma}{\sigma_0} - 2.
\end{align*}
Here, $\mu_0$ and $\sigma_0$ are the parameters of the typical distribution, while $\Delta_\mu$ and $\Delta_\sigma$ denote the strengths of the change in mean and variance respectively.

To simplify the simulations, we assume that the standard deviation remains unaffected by collective anomalies, i.e.\ $\sigma = \sigma_0$. Without loss of generality, we then set $\mu_0=0$ and $\sigma_0 = 1$. Consequently, the strength of the change only depends on the mean $\mu$ of the collective anomaly and is given by
\begin{align}
  \label{eq:strength_sigma}
  \Delta = \log \left( 1 + \frac{\mu^2}{4} \right).
\end{align}
We investigate a number of differing strengths $\Delta = \{0.05,0.1,0.2\}$ corresponding to mean changes of $\mu = \{0.45,0.65,0.94\}$.

In all the simulations reported below we set the minimum segment length to be $l=2$, the maximum segment length to be $m=1000$ and used a burn-in period of $n_0= 1000$ time points. To estimate the ARL, data from the typical regime was simulated and SCAPA ran until the first anomaly was (erroneously) detected. To estimate the ADD, $n_0$ observations were simulated from the typical regime followed by simulated observations from a distribution with an altered mean. We ran SCAPA on this data and calculated the detection delay as being the number of observations after $n_0$ when the anomaly was detected.

\subsubsection{Case 1: IID Gaussian Errors}

\begin{figure}[!h]
  \centering
  \includegraphics[scale=0.4]{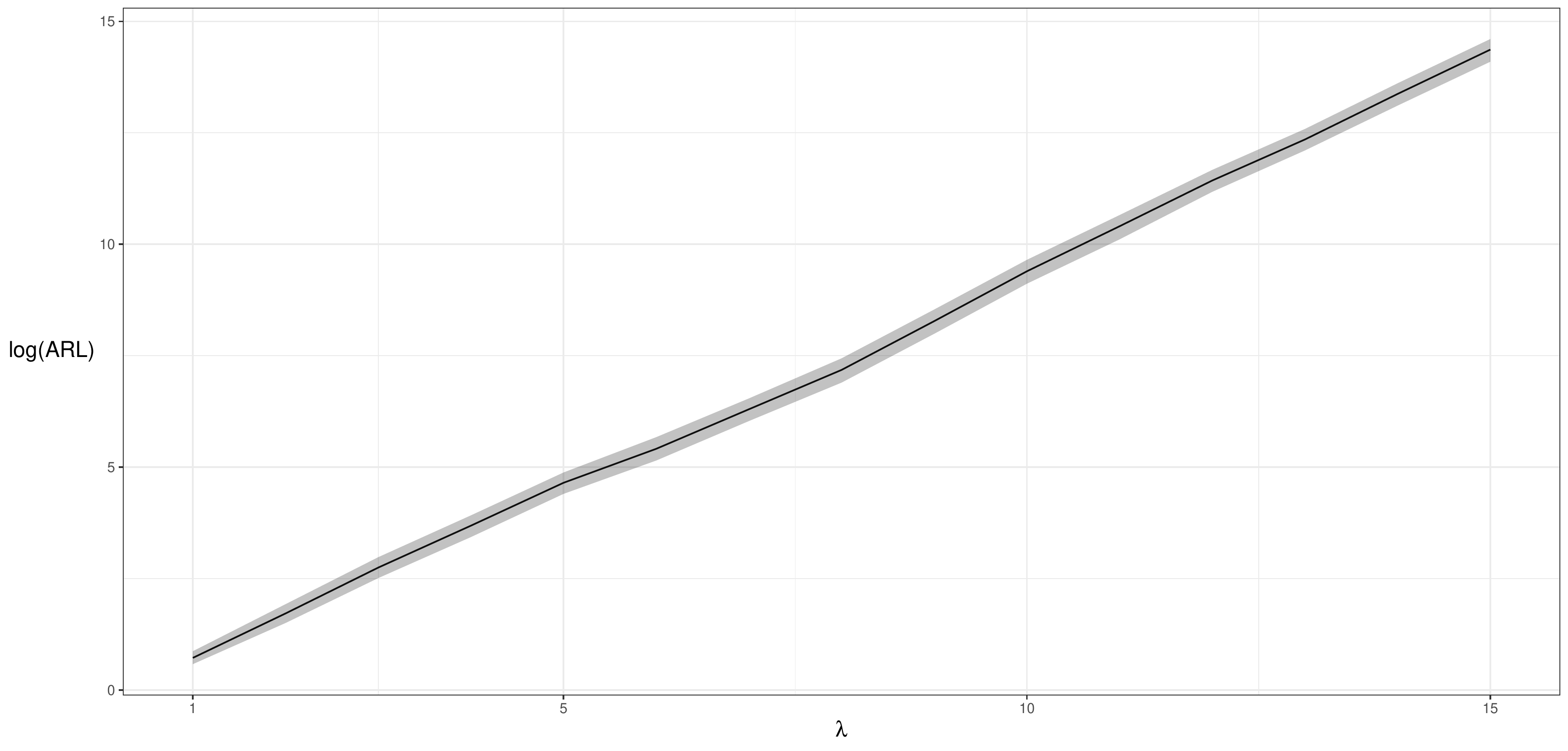}
  \caption{The solid line shows the log-ARL for SCAPA as a function of $\lambda$. The grey shaded region is a pointwise  95\% bootstrapped confidence interval. Results shown from 500 replications.}
  \label{fig:ARL}
\end{figure}

\begin{figure}[h!]
  \centering
  \includegraphics[scale=0.4]{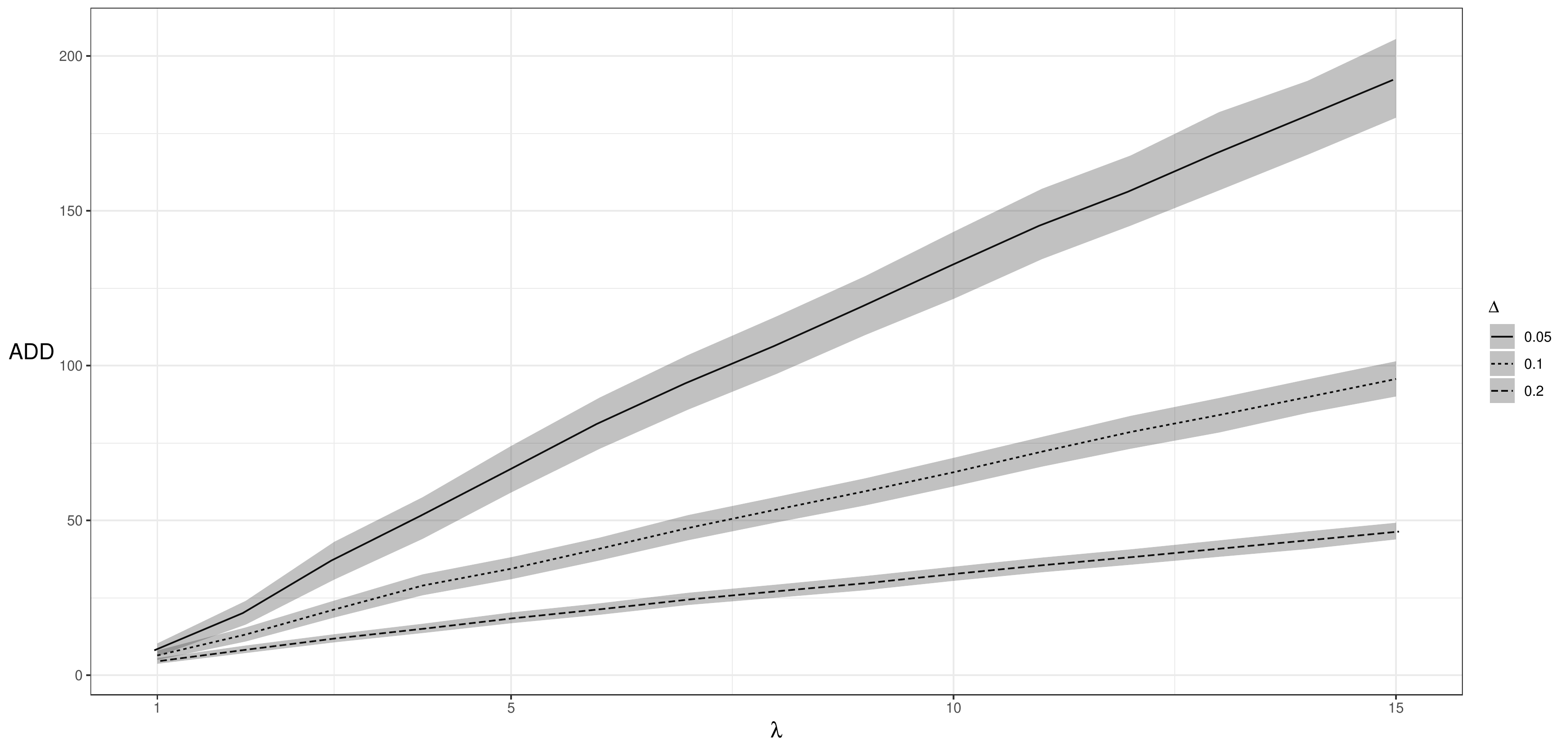}
  \caption{The lines show the ADD for SCAPA as a function of $\lambda$ for different strengths of collective anomaly ($\Delta = 0.05$, 0.1 and 0.2). The grey shaded regions are pointwise 95\% bootstrapped confidence intervals. Results shown from 500 replications.}
  \label{fig:ADD}
\end{figure}

For our initial simulations, we simulated from the assumed model with standard Gaussian errors. Figure \ref{fig:ARL} depicts the log-ARL over a range of values for the penalties \eqref{eq:pens_t} indexed by the parameter $\lambda$ as in \eqref{eq:pens_t} along with a bootstrapped 95\% confidence interval. Similarly, Figure \ref{fig:ADD} shows the relationship between $\lambda$ and the ADD over a range of different values for the mean change of the collective anomaly. 

Note that the log-ARL increases linearly with $\lambda$. 
This structure is reminiscent of the theoretical exponential relationship between $\lambda$ and the ARL derived by \citet{Cao2017RobustSC}, even though these results were derived for known pre and post change behaviour. 

\subsubsection{Case 2: Temporal Dependence}
\label{sec:temporal_dependence}

Whilst the i.i.d.\ data setting is appealing theoretically,
many observed time series are not independent (in time).  Instead many data sequences display serial auto-correlation. To assess the robustness of SCAPA to temporal dependence we simulated an AR(1) error process as the typical distribution, $x_t$, with standard normal errors $\epsilon_t$,
\begin{align*}
  x_t = \phi x_{t-1} + e_t.
\end{align*}
This process was simulated for a range of values of $\phi \in \{0,0.1,0.2,0.3,0.4\}$, representing mild to moderate auto-correlation.

\begin{figure}[h!]
  \centering
  \includegraphics[scale=0.4]{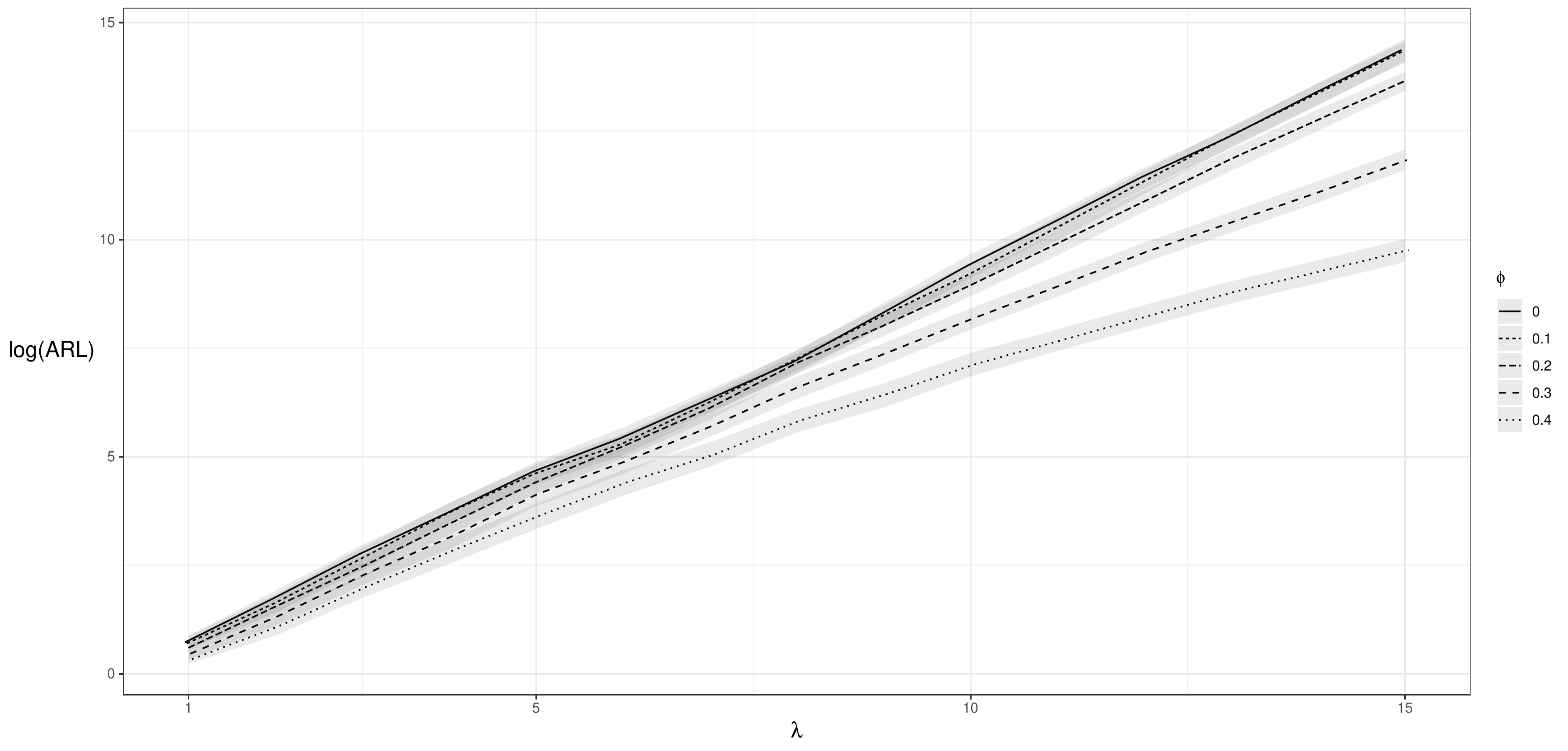}
  \caption{The lines show the log-ARL for SCAPA as a function of $\lambda$ where the simulated time series are AR(1) processes with differing lag-1 auto-correlation ($\phi = 0$, 0.1, 0.2, 0.3 and 0.4). The two penalties, $\beta_C(\lambda)$ and $\beta_O(\lambda)$ are the same as in the i.i.d.\ case (Figure \ref{fig:ARL}). The grey shaded regions are pointwise 95\% bootstrapped confidence intervals. Results shown from 500 replications.}
  \label{fig:depARL}
\end{figure}

As can be seen in Figure \ref{fig:depARL}, the presence of auto-correlation in the residuals leads to the spurious detection of collective anomalies at a higher rate than for independent residuals. This is due to the fact that the cost functions of Section \ref{sec:background} assumed i.i.d data. However, \cite{doi:10.1080/00401706.2018.1438926} gave some empirical evidence that changepoints could be recovered even when auto-correlation is present by applying a correction or inflation factor to the penalty. This factor is the sum of the auto-correlation function for the residuals from $-\infty$ to $\infty$. This is equal to $(1+\phi)/(1-\phi)$ for the AR(1) model. A similar correction exists for MA processes. We repeated the simulations using this correction. 

\begin{figure}[h!]
  \begin{subfigure}[t]{0.3\textwidth}
    \includegraphics[width=\textwidth]{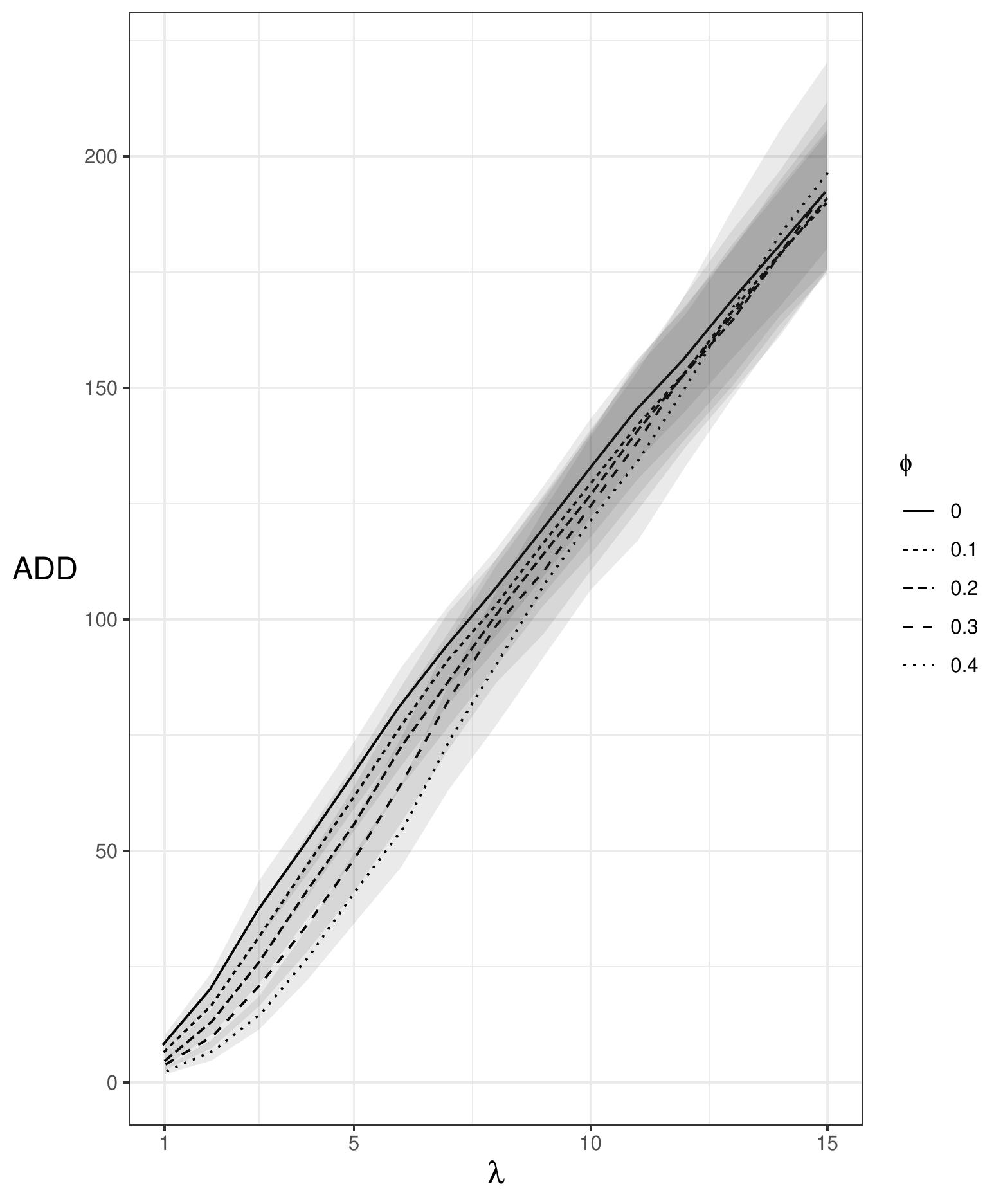}
    \caption{} 
  \end{subfigure}\hfill
  \begin{subfigure}[t]{0.3\textwidth}
    \includegraphics[width=\textwidth]{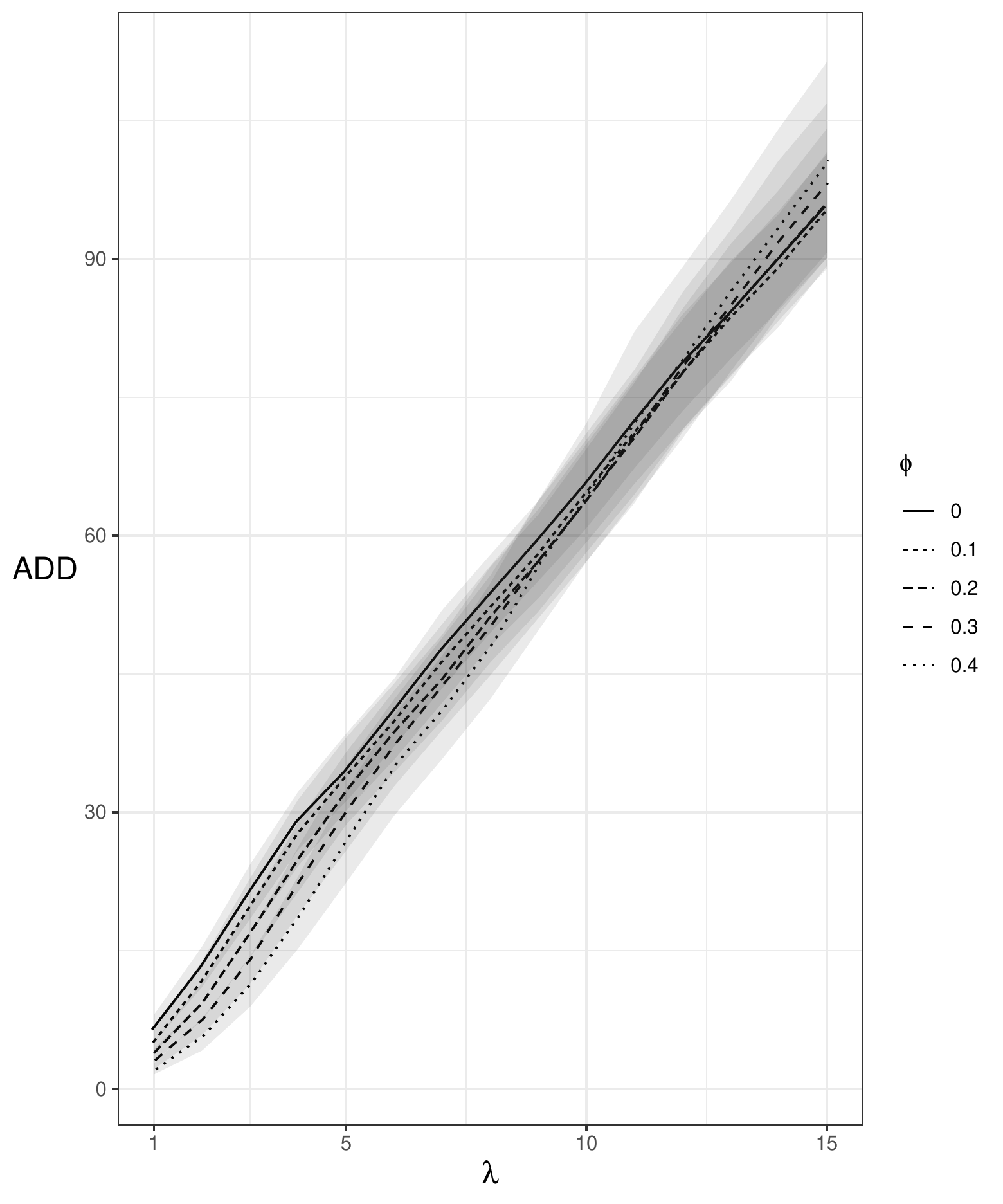}
    \caption{} 
  \end{subfigure}
  \hfill
  \begin{subfigure}[t]{0.3\textwidth}
    \includegraphics[width=\textwidth]{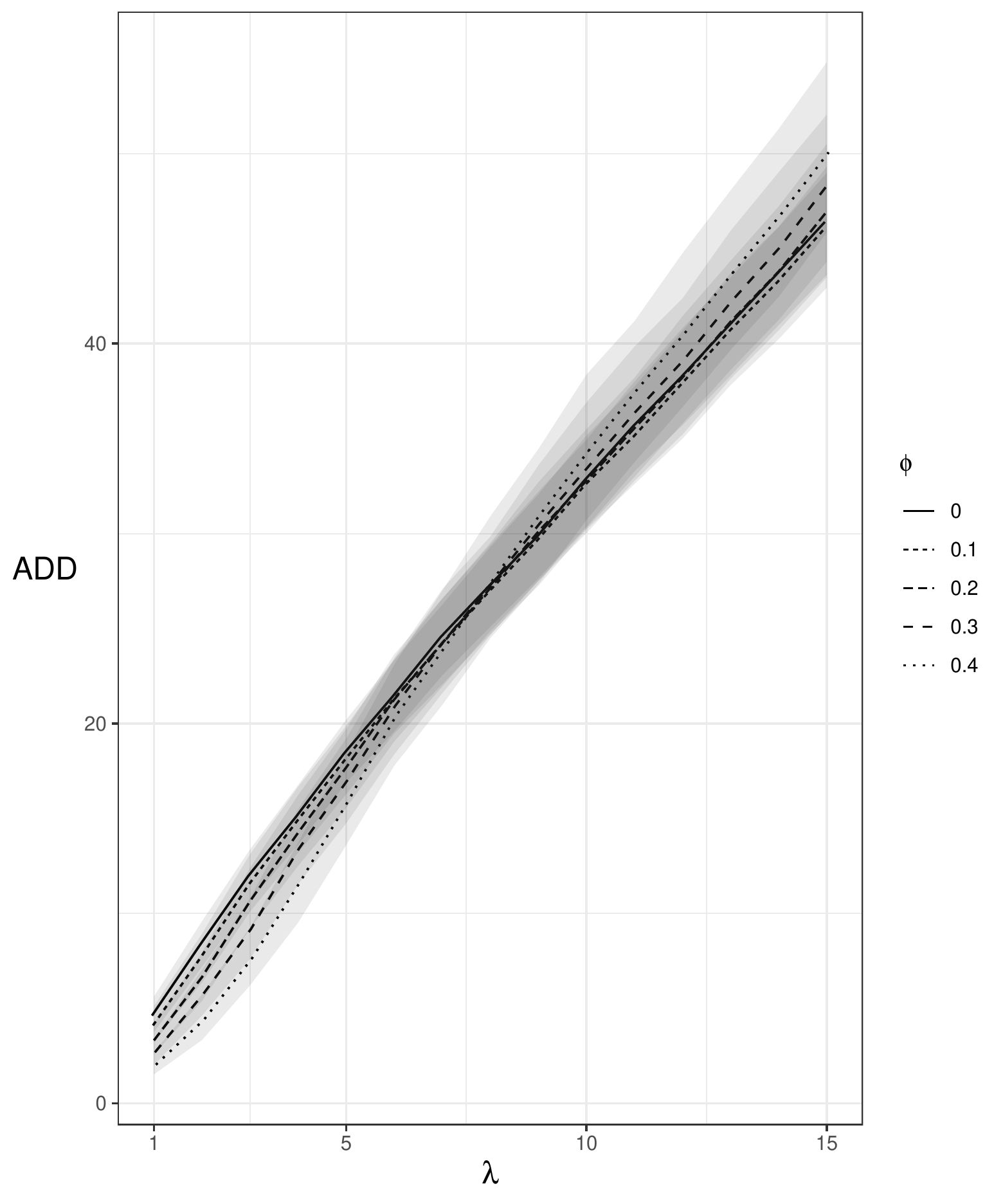}
    \caption{} 
  \end{subfigure}
   \caption{The lines show the ADD for SCAPA as a function of $\lambda$ for different strengths of collective anomaly a) $\Delta = 0.05$, b) $\Delta = 0.1$ and c) $\Delta = 0.2$. In each case the simulated residuals are AR(1) processes with differing lag-1 auto-correlation ($\phi = 0$, 0.1, 0.2, 0.3 and 0.4). The two penalties, $\beta_C(\lambda)$ and $\beta_O(\lambda)$ are the same as in the i.i.d.\ case (Figure \ref{fig:ADD}). The grey shaded regions are pointwise 95\% bootstrapped confidence intervals. Results shown from 500 replications.}
  \label{fig:depADD}
\end{figure}

\begin{figure}[h!]
  \centering
  \includegraphics[scale=0.4]{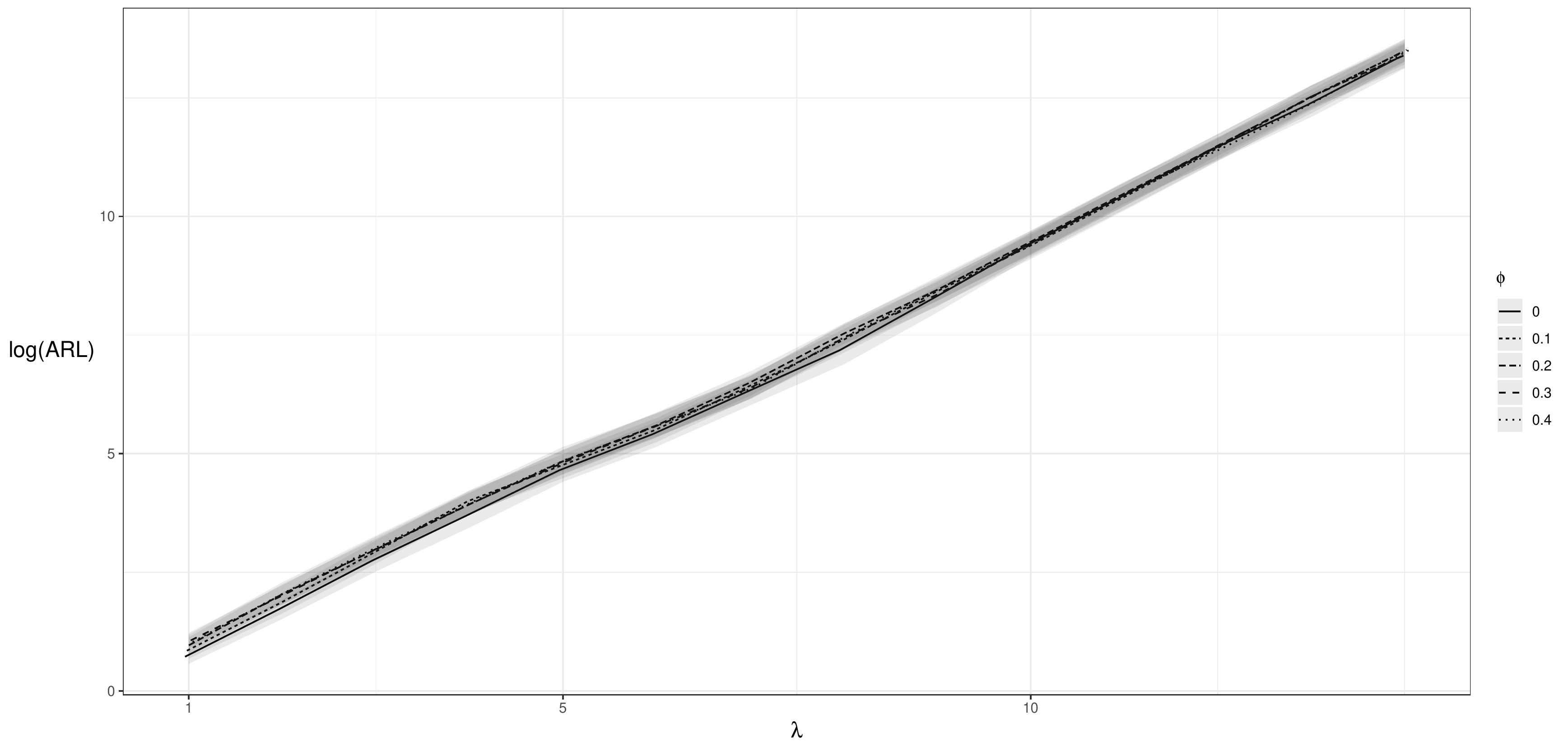}
  \caption{The lines show the log-ARL for SCAPA as a function of $\lambda$ where the simulated time series are AR(1) processes with differing lag-1 auto-correlation ($\phi = 0$, 0.1, 0.2, 0.3 and 0.4). The two penalties, $\beta_C(\lambda)$ and $\beta_O(\lambda)$ are inflated by a function of $\phi$. The grey shaded regions are pointwise 95\% bootstrapped confidence intervals. Results shown from 500 replications.}
  \label{fig:dep_corrected_ARL}
\end{figure}

\begin{figure}[h!]
  \begin{subfigure}[t]{0.3\textwidth}
    \includegraphics[width=\textwidth]{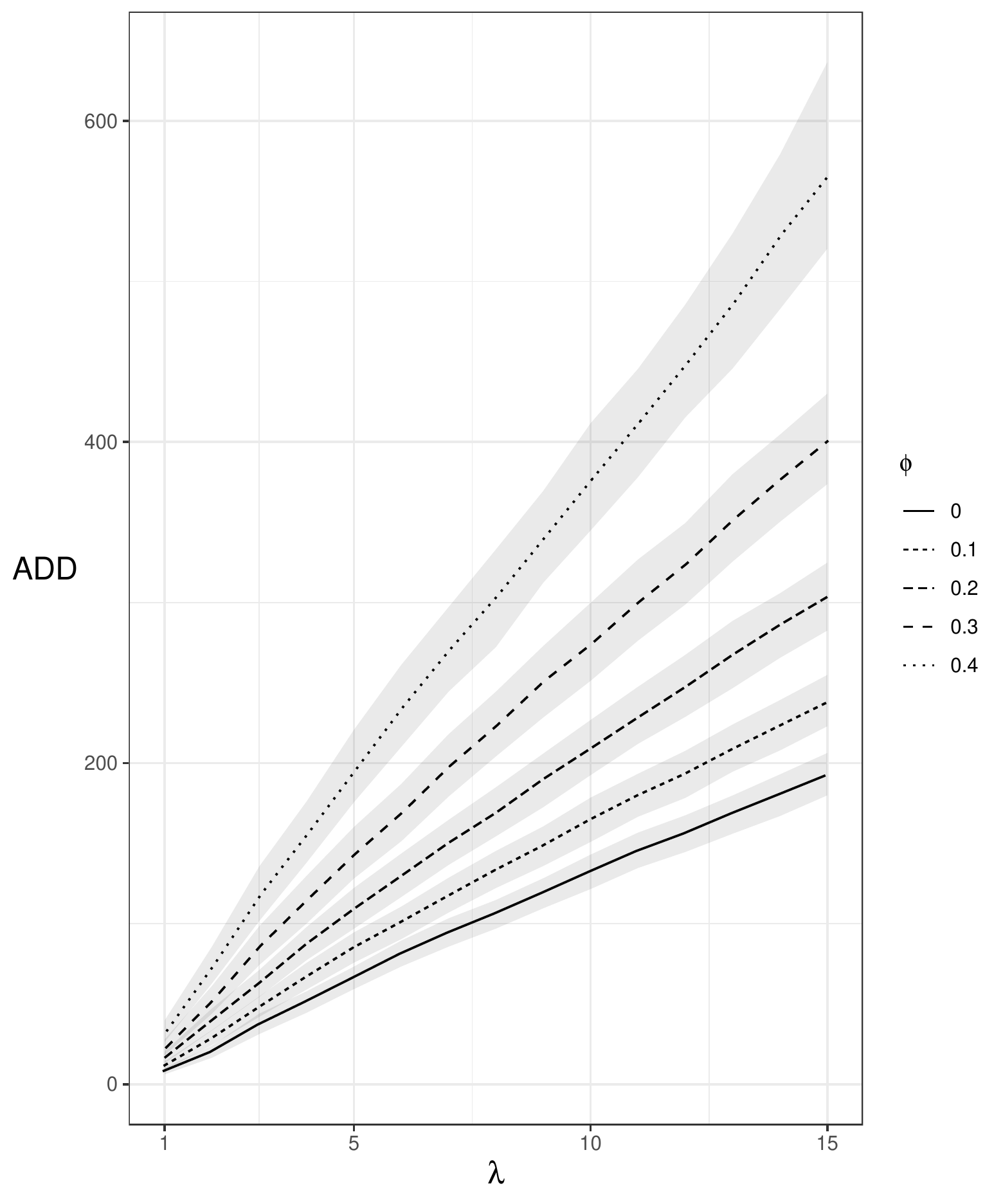}
    \caption{}
  \end{subfigure}\hfill
  \begin{subfigure}[t]{0.3\textwidth}
    \includegraphics[width=\textwidth]{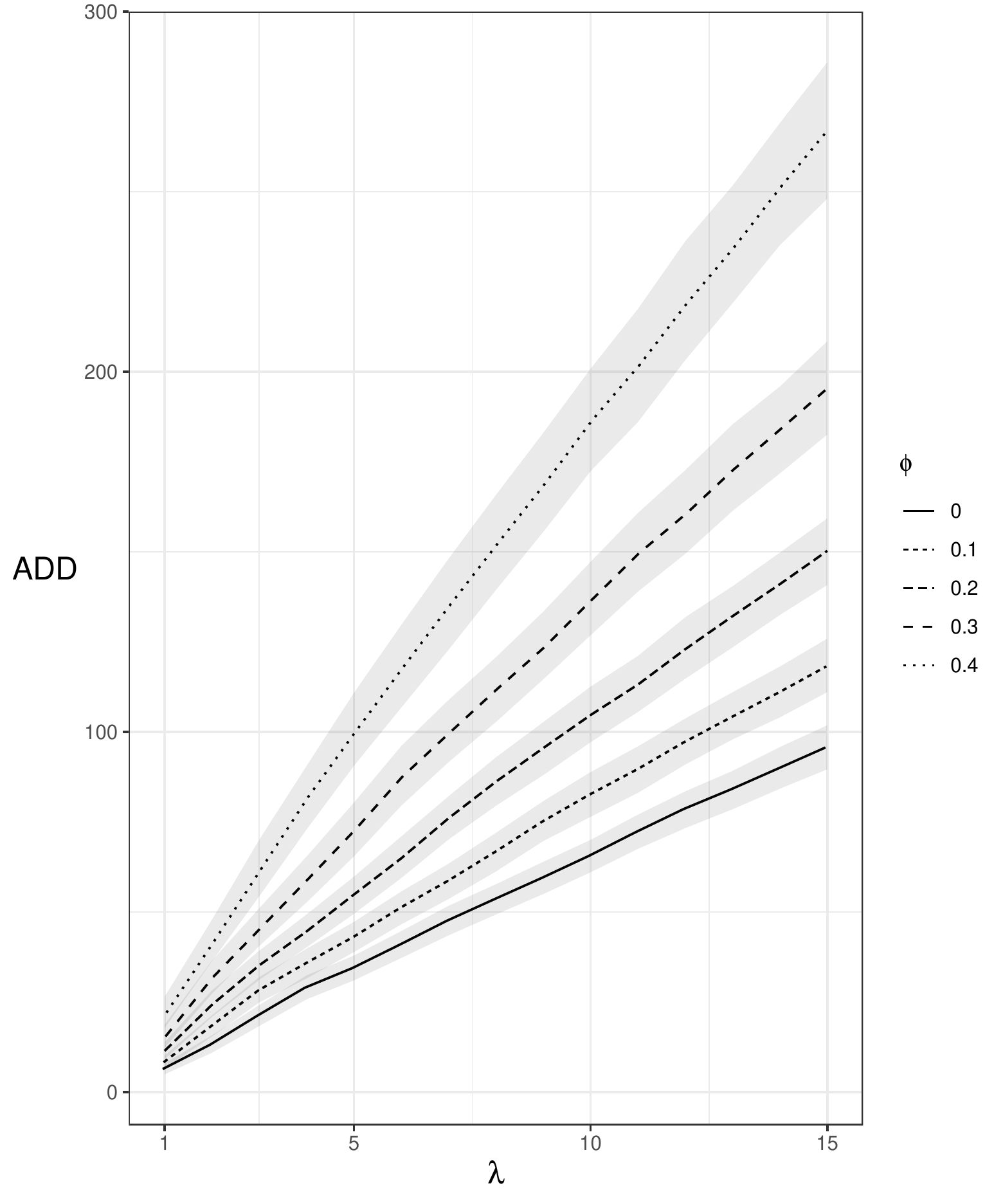}
    \caption{}
  \end{subfigure}
  \hfill
  \begin{subfigure}[t]{0.3\textwidth}
    \includegraphics[width=\textwidth]{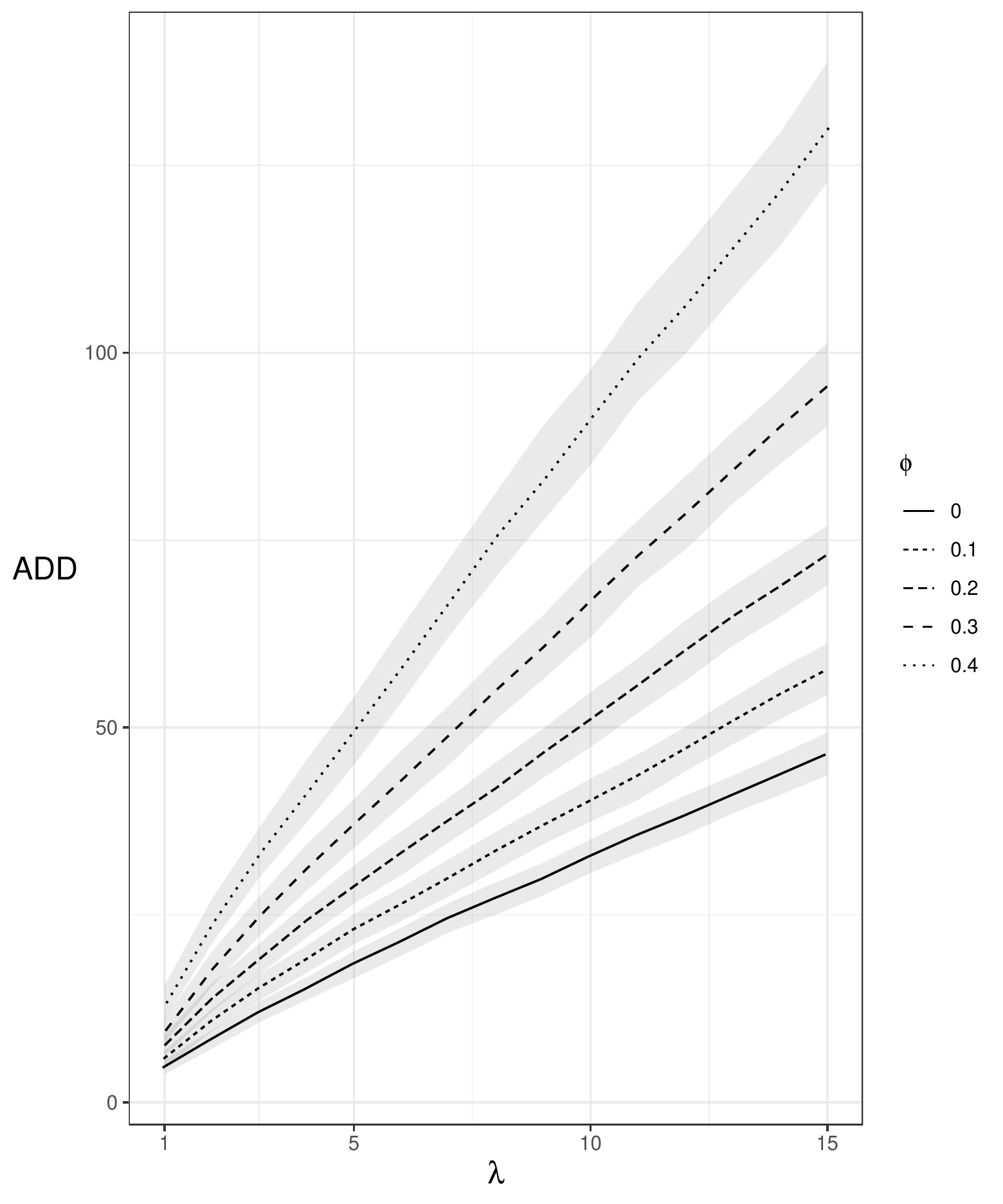}
    \caption{}
  \end{subfigure}
   \caption{The lines show the ADD for SCAPA as a function of $\lambda$ for different strengths of collective anomaly a) $\Delta = 0.05$, b) $\Delta = 0.1$ and c) $\Delta = 0.2$. In each case the simulated residuals are AR(1) processes with differing lag-1 auto-correlation ($\phi = 0$, 0.1, 0.2, 0.3 and 0.4). The two penalties, $\beta_C(\lambda)$ and $\beta_O(\lambda)$ are inflated by a function of $\phi$. The grey shaded regions are pointwise 95\% bootstrapped confidence intervals. Results shown from 500 replications.}
  \label{fig:dep_corrected_ADD}
\end{figure}

 The results in Figure \ref{fig:dep_corrected_ARL} show that the log-ARL of SCAPA with appropriately inflated penalties is almost identical to that of the i.i.d.\ case. On the other hand, the ADD now depends on the auto-correlation due to the inflated penalty (see Figure \ref{fig:dep_corrected_ADD}).

Performing this correction requires knowledge of the AR(1) parameter $\phi$. If it is unknown it can be estimated robustly either on a batch of the data or sequentially using SA-estimates (see \citep{Sharia2010})

\subsection{Multiple Anomalies} \label{sec:Multi}

A natural comparison to make when investigating an online method is to compare its performance to its offline counterpart. We therefore compare SCAPA and CAPA for the detection of multiple anomalies using ROC curves in this section. 

To this end, we simulated time series with a total length of 10,000 observations with a number of point and collective anomalies. The length of stay for the typical state and for collective anomalies were sampled from a $\textrm{NB}(5,0.01)$ distribution and a $\textrm{NB}(5,0.03)$ distribution respectively. Observations in the typical state were sampled from an $N(0,1)$ distribution, while observations from the $k$th collective anomaly were sampled from an $N(\mu_k,\sigma_k^2)$ distribution, where $\mu_1,...,\mu_K \sim N(0,2^2)$ and $\sigma_1,...\sigma_K \sim \Gamma(1,1)$. Point anomalies occurred in the typical state independently with probability $p = 0.01$ and were drawn from a $t$-distribution with 2 degrees of freedom.

\begin{figure}[h!]
  \centering
  \includegraphics[scale=0.4]{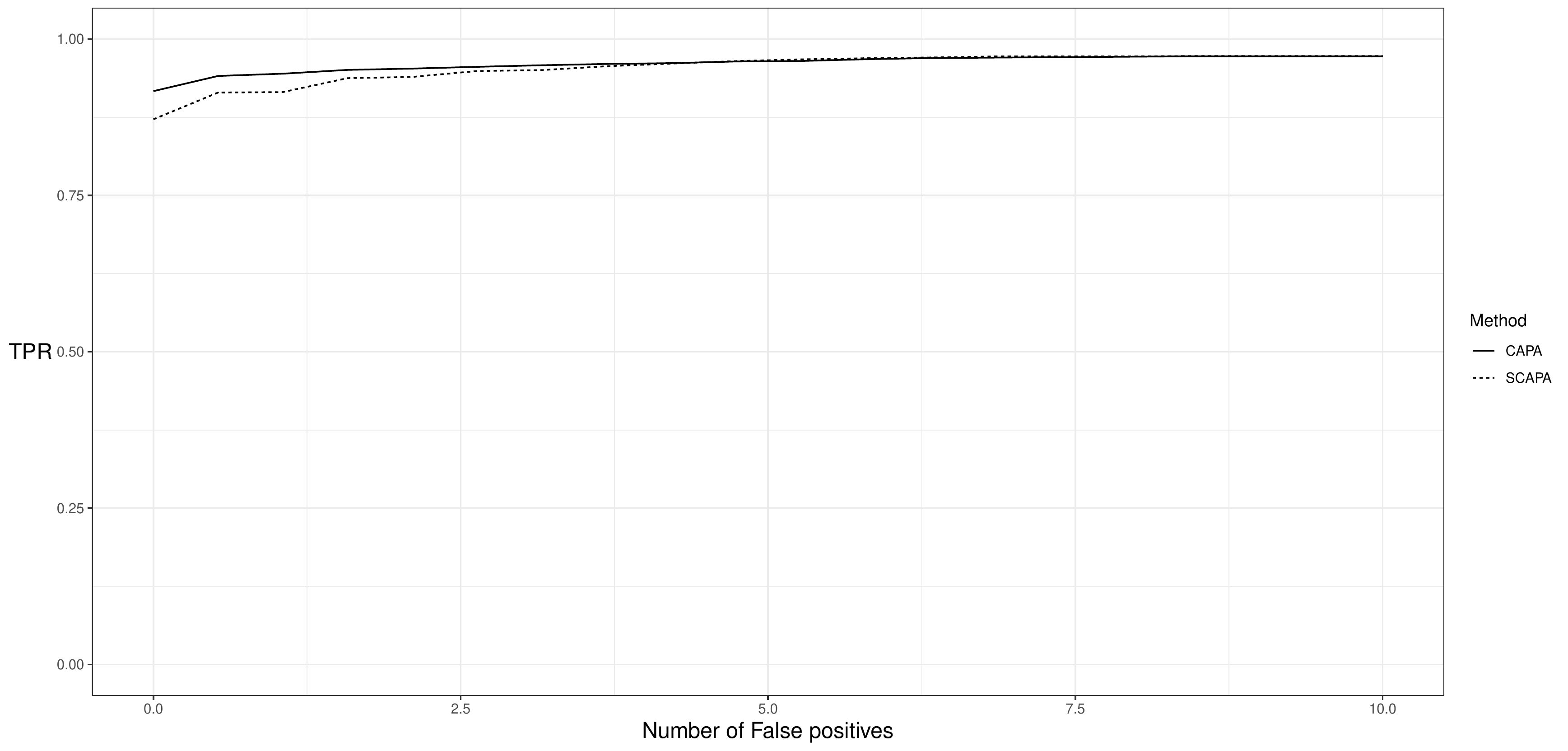}
  \caption{ROC curves for CAPA (solid line) and SCAPA (dashed line)
  from 100 replications.}
  \label{fig:ROC}
\end{figure}

\begin{figure}[h!]
  \begin{subfigure}[t]{0.5\textwidth}
    \centering
    \includegraphics[scale=.3]{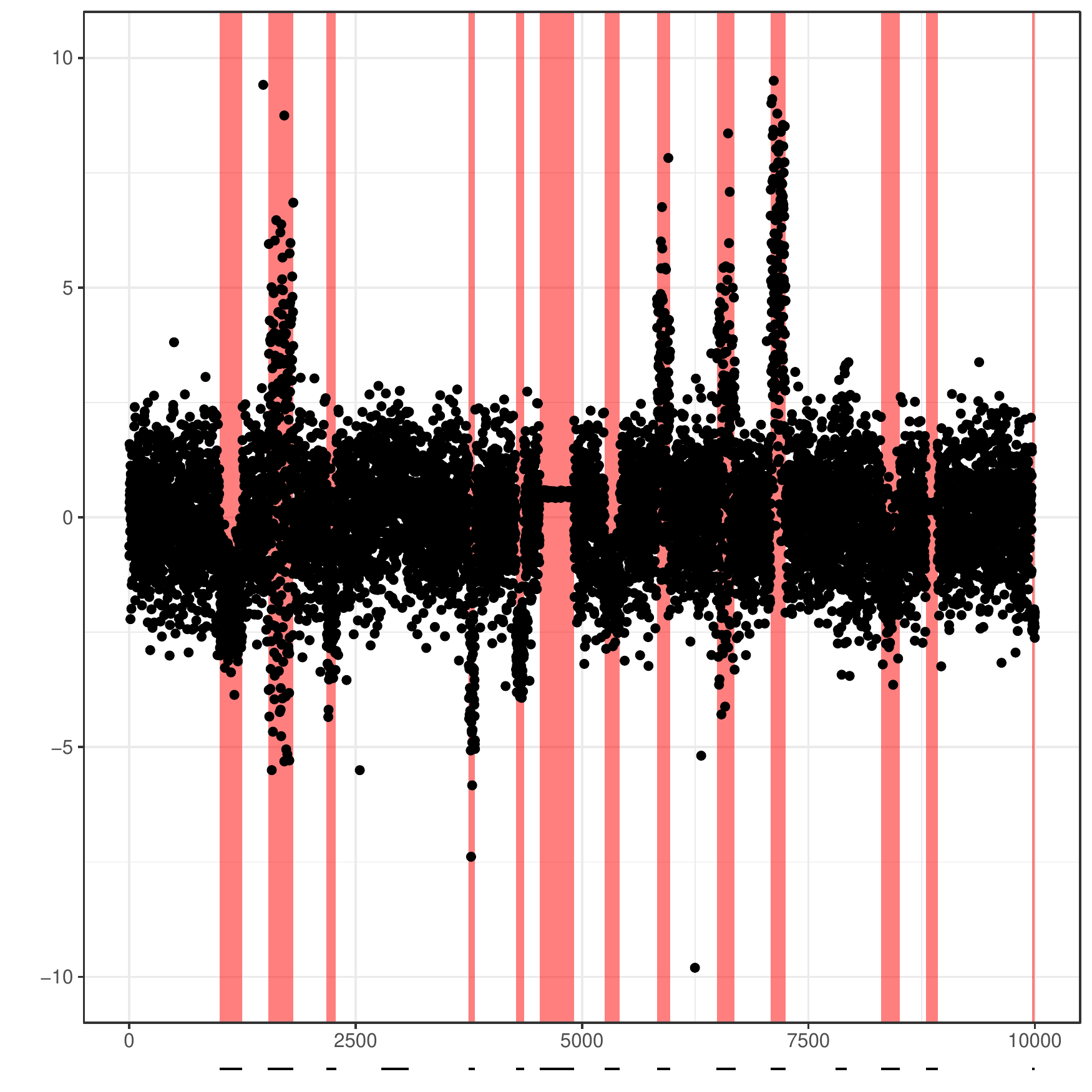}
    \caption{}
  \end{subfigure}\hfill
  \begin{subfigure}[t]{0.5\textwidth}
    \centering
    \includegraphics[scale=.3]{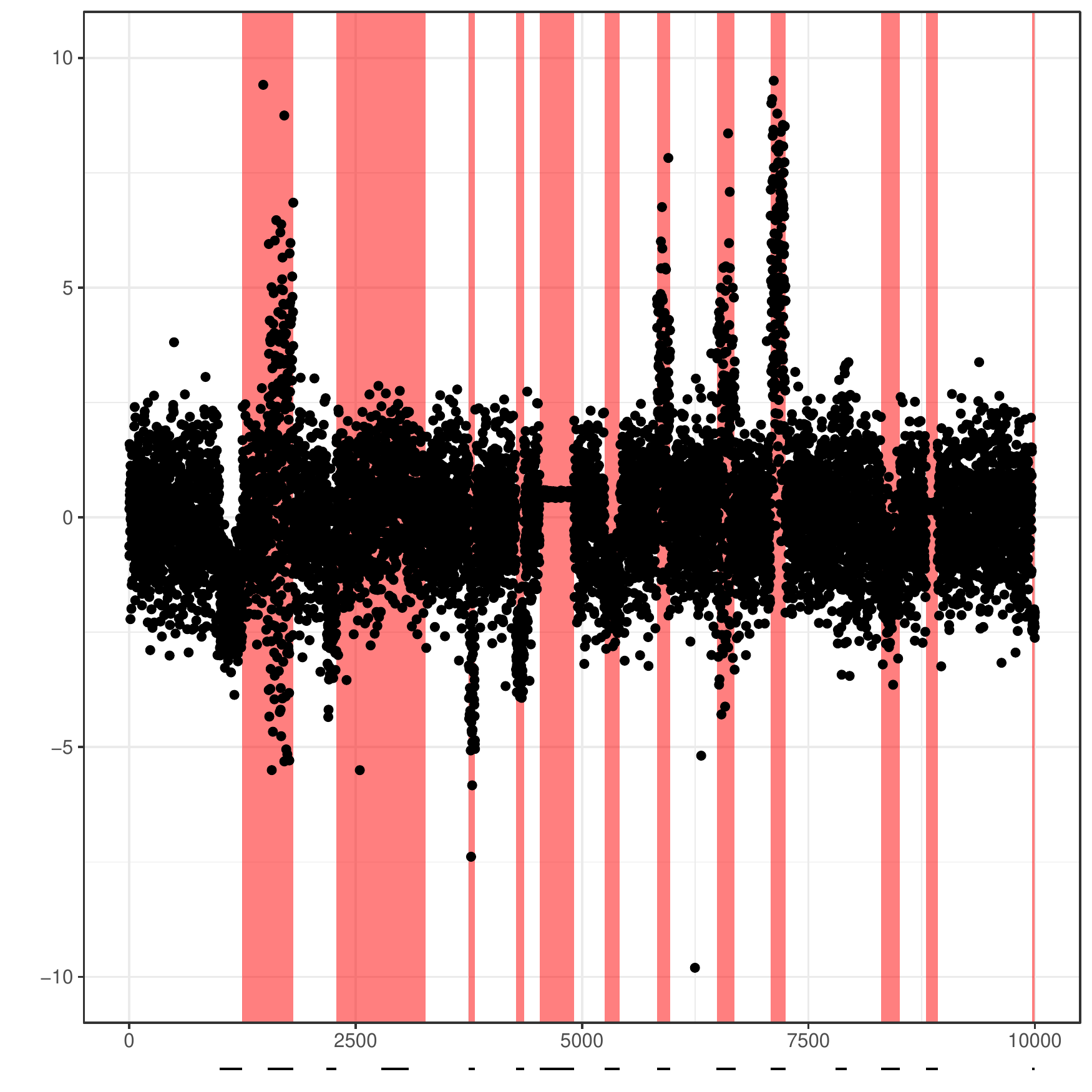}
    \caption{}
  \end{subfigure}
\caption{A comparison of a) CAPA to b) SCAPA on an example time series. Segments in red show inferred collective anomalies. Dashed lines below the $x$-axis show the position of the true collective anomalies in the data.}
\label{fig:SCAPA_ROC_ts}
\end{figure}

The ROC curve resulting from this simulation can be found in Figure \ref{fig:ROC}, alongside an example time series in Figure \ref{fig:SCAPA_ROC_ts} shown segmented by both CAPA and SCAPA. As expected, CAPA, which has access to the whole data outperforms SCAPA which tends to overestimate the number of anomalies. However, the gap is small, especially for low values of $\lambda$. 

\subsection{CUSUM comparison} \label{sec:CUSUM}

A natural comparison that can be made to assess SCAPA's simulated performance is with the widely used online change point detection method CUSUM \citep{citeulike:3720621}. Both methods use a test statistic based on the log-likelihood ratio and can be configured with known typical mean and variance. The difference between the two methods is that in SCAPA, collective and point anomalies are detected jointly with separate penalties whereas CUSUM is not designed to be robust to point anomalies. In our simulations, the CUSUM approach is implemented by setting the penalty for point anomalies in SCAPA to an arbitrarily large value ($\beta_O = 10^{12}$) so that no point anomalies are detected.

\begin{figure}[h!]
  \begin{subfigure}[t]{0.5\textwidth}
    \centering
    \includegraphics[scale=.3]{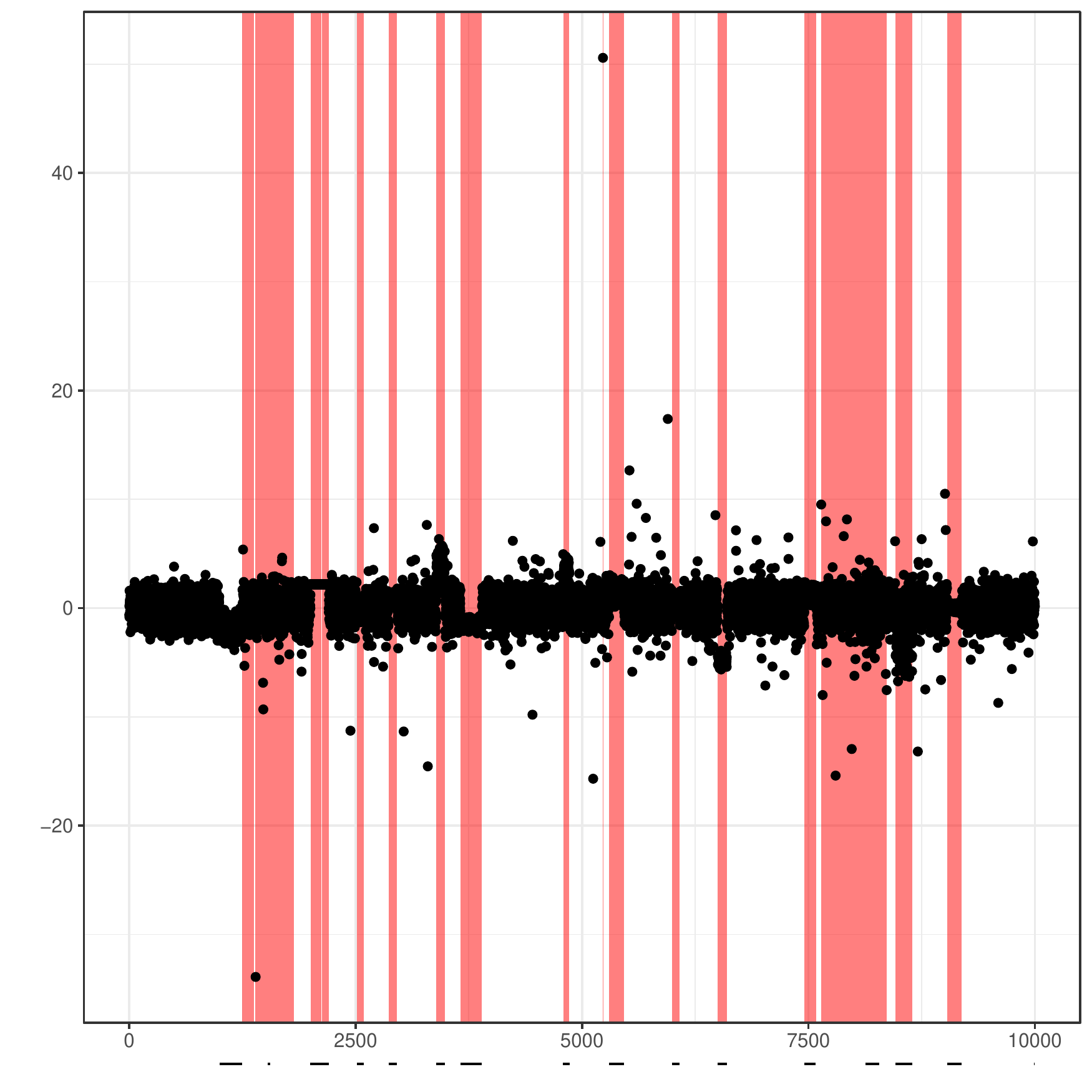}
    \caption{}
  \end{subfigure}\hfill
  \begin{subfigure}[t]{0.5\textwidth}
    \centering
    \includegraphics[scale=.3]{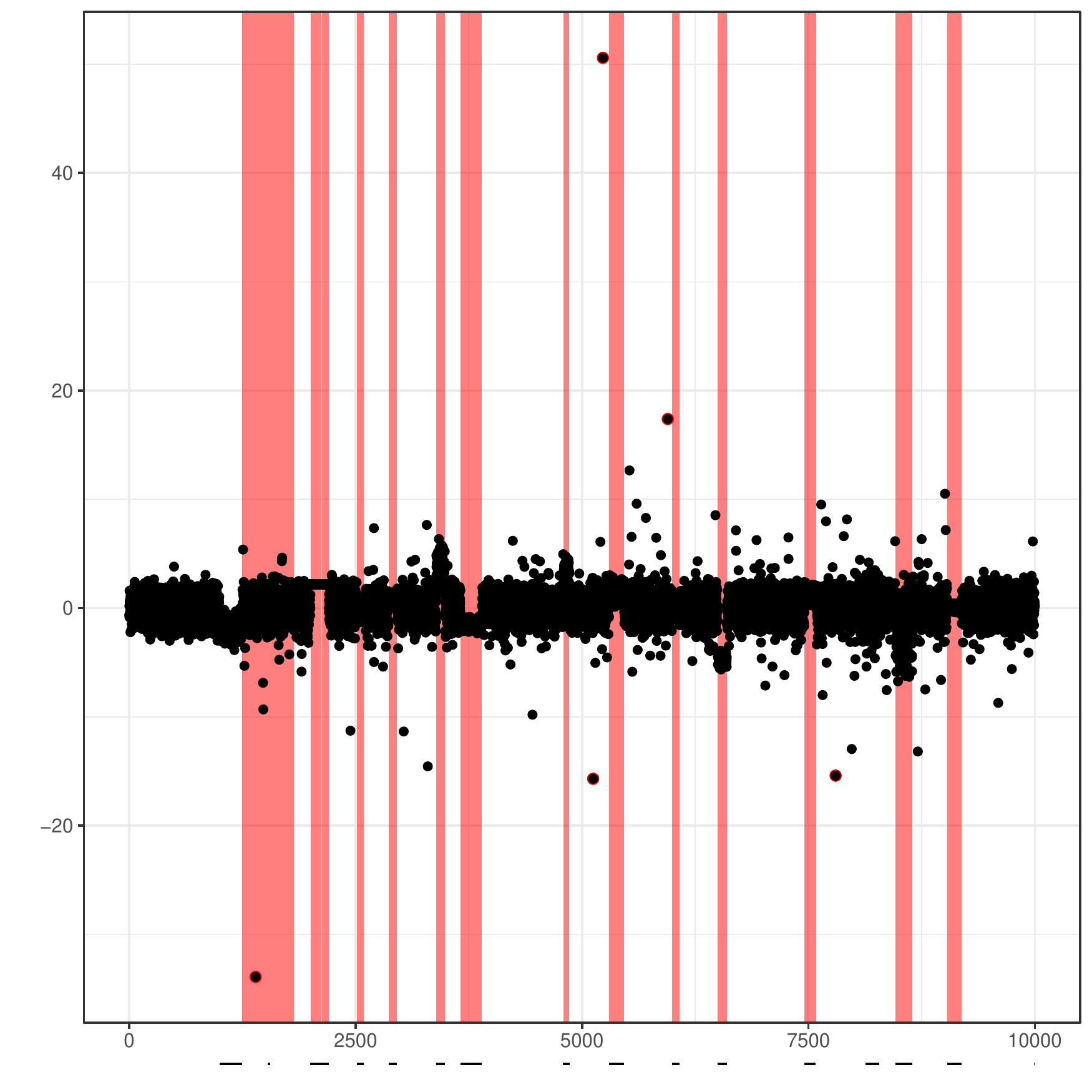}
    \caption{}
  \end{subfigure}
\caption{A comparison of a) CUSUM to b) SCAPA on an example time series. Segments in red show inferred collective anomalies. Dashed lines below the $x$-axis show the position of the true collective anomalies in the data.}
\label{fig:SCAPA_CUSUM}
\end{figure}

The data was simulated in a similar way to that of Section \ref{sec:Multi} with the difference being that 20\% of points in the typical state were point anomalies,  simulated from a $t$-distribution with degree of freedom $\nu \in \{2,5,10\}$. 
\begin{figure}[h!]
  \begin{subfigure}[t]{0.5\textwidth}
    \centering
    \includegraphics[scale=.4]{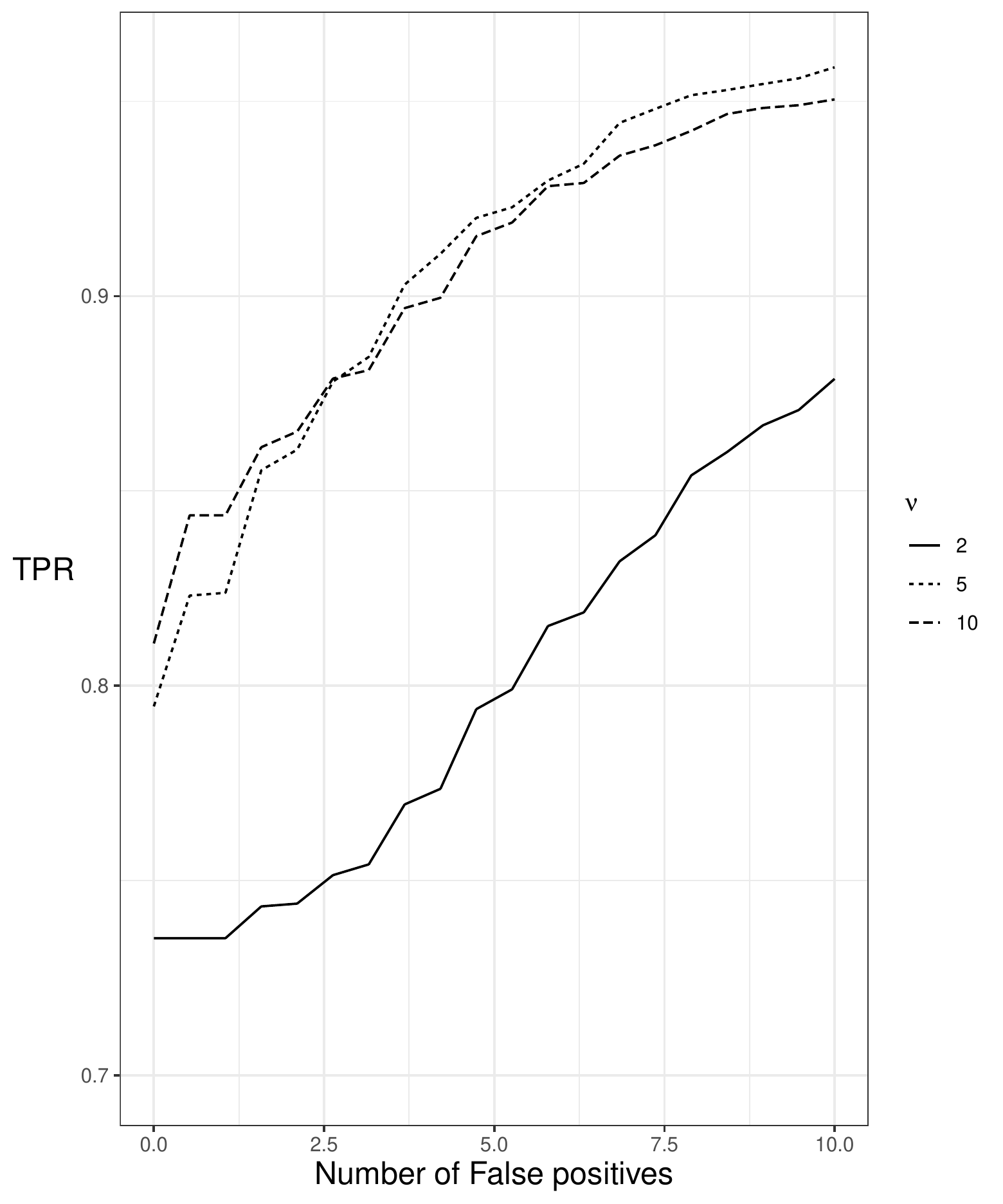}
    \caption{}
  \end{subfigure}\hfill
  \begin{subfigure}[t]{0.5\textwidth}
    \centering
    \includegraphics[scale=.4]{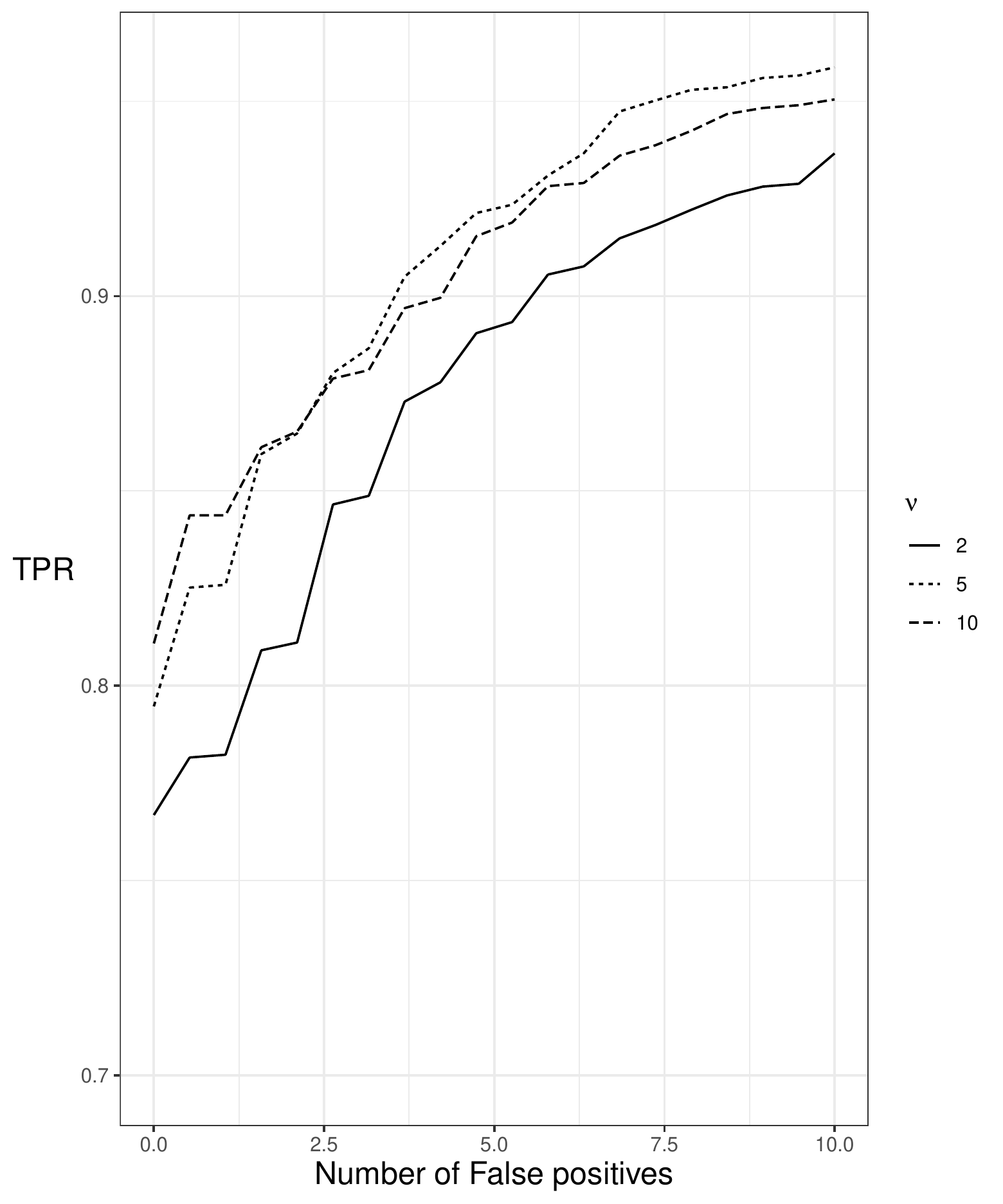}
    \caption{}
  \end{subfigure}
\caption{ROC curves over 100 replications for a) CUSUM and b) SCAPA. Point anomalies were generated from a $t$-distribution with varying degrees of freedom ($\nu$ = 2, 5 or 10 respectively).}
\label{fig:ROC_CUSUM}
\end{figure}
The ROC curve resulting from this simulation can be found in Figure \ref{fig:ROC_CUSUM}, alongside an example time series in Figure \ref{fig:SCAPA_CUSUM} shown segmented by both methods. As expected CUSUM gives a higher number of false positives than SCAPA when point anomalies are from distributions with heavier tails.

%% file: sections/machine_temp.tex
\section{Machine Temperature Data}
\label{sec:temp}

The Numenta Anomaly Benchmark (NAB) \citep{Lavin2015EvaluatingRA,AHMAD2017134} provides a number of data sets that can be used to compare different anomaly detection approaches. The data can be obtained from \url{https://github.com/numenta/NAB}. 

\begin{figure}[h]
  \centering
  \includegraphics[scale=0.4]{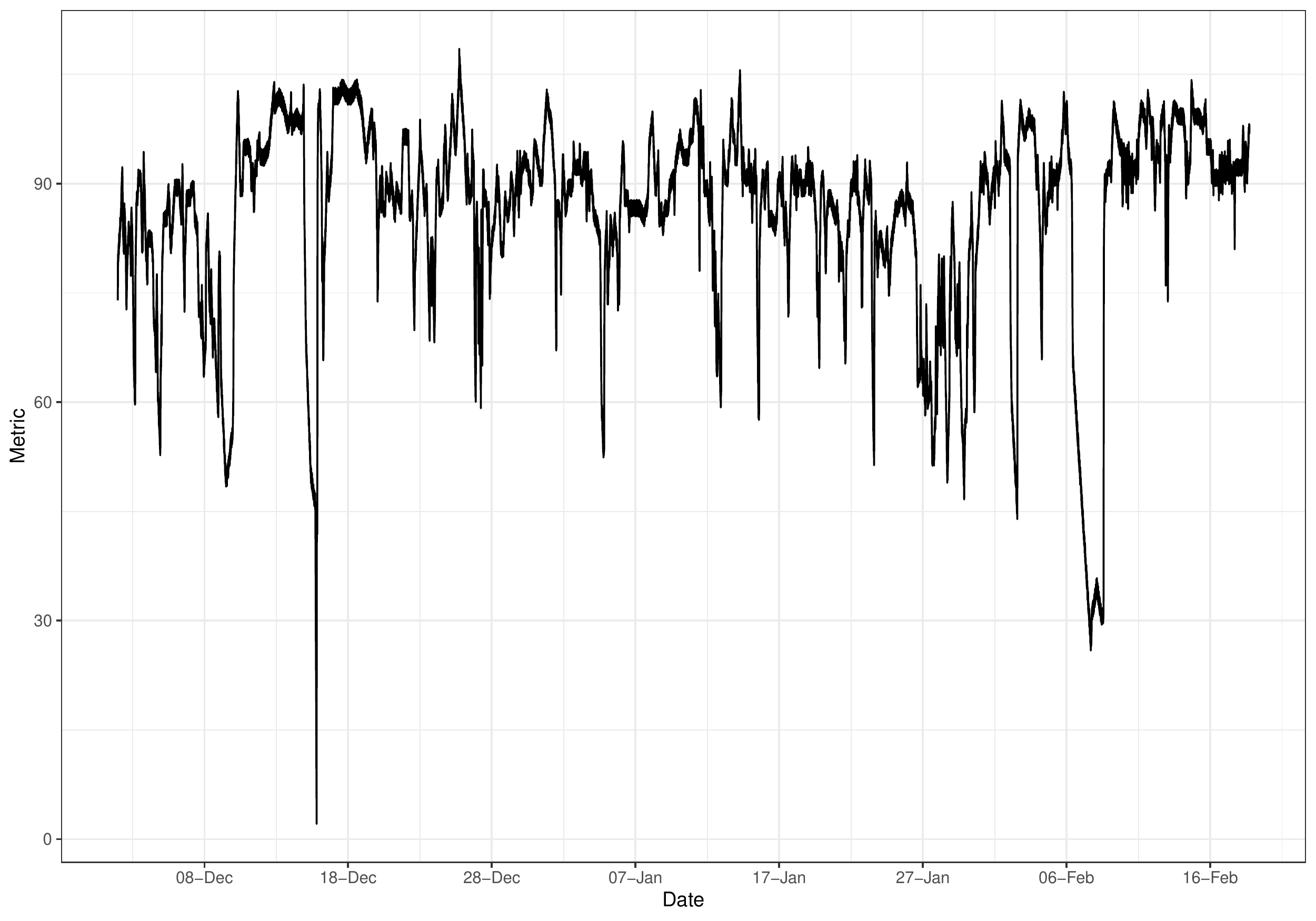}
  \caption{Machine temperature data}
  \label{fig:machine_temp}
\end{figure}

One example consists of heat sensor data from an internal component of a large industrial machine. The data is displayed in Figure \ref{fig:machine_temp}. There are $n = 22,695$ observations spanning 2nd December 2013 - 19th February 2014 sampled every five minutes. 

\begin{figure}[h]
  \centering
  \includegraphics[scale=0.4]{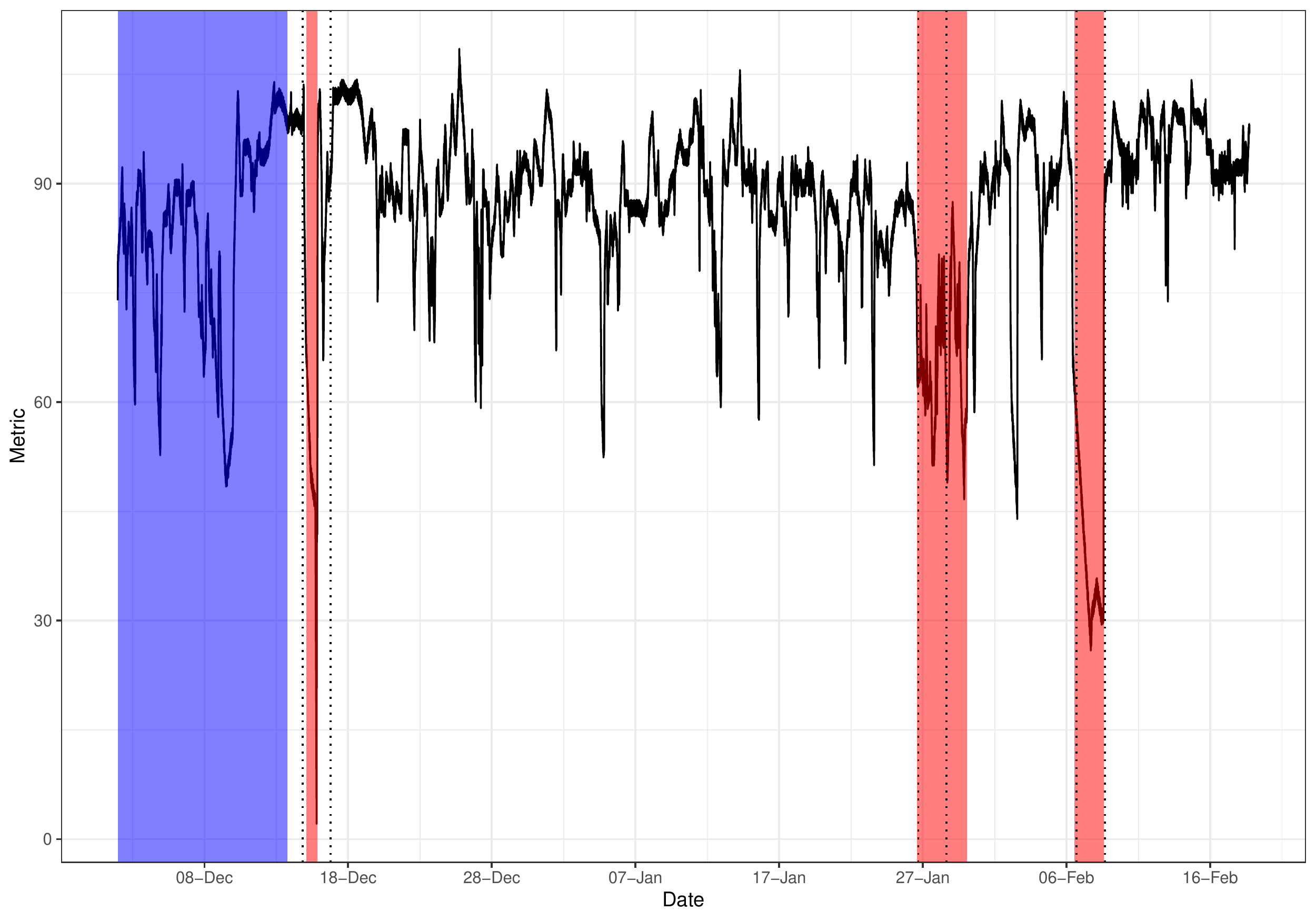}
  \caption{Machine temperature data. Detected anomalies are shaded in red and the burn-in period in blue. Dashed vertical lines show the hand labelled anomalies given by an engineer working on the machine.}
  \label{fig:machine_temp_anom}
\end{figure}

\cite{Lavin2015EvaluatingRA} use an initial, or burn-in, period to allow their algorithms to learn about the data. In line with their approach, we set the burn in period to be the first 15\% of the data (2nd December 2013 until the 14th December 2013, as shown by the blue shaded area in Figure \ref{fig:machine_temp_anom}). We used the burn in to obtain a robust M-estimator for the lag-1 autocorrelation of the observations after standardisation by the sequential mean and variance estimate. Using the robust $M$ estimator of \cite{10.2307/2242597} from the \texttt{R} package \texttt{robust} \citep{robustpackage}, we obtained an autocorrelation estimate $\hat{\phi}=0.974$. In line with the approach taken in Section \ref{sec:temporal_dependence}, we therefore set the penalties to:
\begin{align*}
  \beta_C = 2 \times \frac{1 + \hat{\phi} }{1 - \hat{\phi} } \times \log (n), \;\;\;\;\;\;\;\;\;
  \beta_O = 2 \times \frac{1 + \hat{\phi} }{1 - \hat{\phi} } \times \log (n).
\end{align*}

Figure \ref{fig:machine_temp_anom} shows the three anomalies SCAPA detected shaded in red. These corresponded to a set of hand labelled anomalous regions given by an engineer working on the machine shown by the dashed vertical lines. The positions of these are given in Table \ref{tab:known_anoms_machine_temp}. It should be noted that the data labels in the NAB consist of anomalous periods, rather than points. However, all approaches previously applied to the data only return points of anomalous behaviour, highlighting SCAPA's potential to provide new insights into the data.

\begin{table}[ht]
  \centering
  \footnotesize
  \begin{tabular}{ccccc}
    \hline\hline
    Anomaly  & Start time       & End time       & Given reason                 & \textbf{Detection time} \\ 
    \hline
    1      & 17:50 15/12/2013 & 17:00 17/12/2013 & Planned shutdown             & \textbf{16:50 16/12/2013}  \\ 
    2      & 14:20 27/01/2014 & 13:30 29/01/2014 & Onset of problem             & \textbf{21:25 28/01/2014}  \\ 
    3      & 14:55 07/02/2014 & 14:05 09/02/2014 & Catastrophic system failure  & \textbf{3:15 08/02/2014}   \\ 
    \hline\hline
  \end{tabular}
	\normalsize
  \caption{Labelled anomalies from the NAB obtained from \url{https://github.com/numenta/NAB/blob/master/labels/combined_windows.json} along with the time it was detected (in bold).}
  \label{tab:known_anoms_machine_temp}
\end{table}

The detection of the more subtle second anomaly in a timely fashion is important as this was claimed in the NAB literature to be the cause of the catastrophic system failure (third anomaly). We can see that the time at which SCAPA first detected it in Table \ref{tab:known_anoms_machine_temp}. If users of the system deemed this to be too long of a delay the penalties used above could be decreased, however, as noted elsewhere in this paper this would increase the frequency of false alarms.

%% file: sections/Acknowledgements.tex
\section{Acknowledgements}

This work was supported by EPSRC grant numbers EP/N031938/1 (StatScale), EP/R004935/1 (NG-CDI) and EP/L015692/1 (STOR-i). Fisch also gratefully acknowledges EPSRC (EP/S515127/1) and British Telecommunications plc (BT) for providing financial support for his PhD via an Industrial CASE award. Finally, the authors thank David Yearling, Trevor Burbridge, Stephen Cassidy, and Kjeld Jensen in BT Research for helpful discussions while this work was being undertaken.

%% file: sections/Supplementary.tex
\section{Supplementary Material}

\subsection{Pseudocode}

\begin{algorithm}
	\caption{Sequential quantile estimation}
	\label{alg:SQE}
	\begin{algorithmic}[1]
		\Function{InitialQuantile}{$\mathbf{x}$, $\alpha$}
		\State $M \gets |\mathbf{x}|$
		\State $\xi \gets \mathbf{x}_{(\alpha)}$
		\State $d_0 \gets \frac{1}{ \mathbf{x}_{(0.75)} - \mathbf{x}_{(0.25)}  }$
		\State $c = \frac{d_0}{M} \sum_{i=1}^{M} i^{-1/2}$
		\State $f = \frac{1}{2cM} \max \left\{ \# \left\{ |\mathbf{x} - \xi| \leq c \right\}, 1 \right\}$
		\State $\textrm{state}.\xi \gets \xi$
		\State $\textrm{state}.d \gets d_{0}$
		\State $\textrm{state}.\hat{f} \gets f$
		\State $\textrm{state}.i \gets 0$
		\State\Return \textrm{state} 
		\EndFunction
		
		\Function{UpdateQuantile}{$\textrm{state}$, $x$, $\alpha$}
		\State $a=1/4$
		\State $\xi \gets \textrm{state}.\xi$
		\State $d \gets \textrm{state}.d$
		\State $\hat{f} \gets \textrm{state}.\hat{f}$
		\State $i \gets \textrm{state}.i$
		\State $\xi = \xi - \frac{d}{i+1} \left( \mathbbm{1}\left[x \leq \xi \right] - \alpha \right) $
		\State $\hat{f} = \frac{1}{i+1} \left( i\hat{f} + \frac{\sqrt{i+1}}{2} \mathbbm{1}\left[ \lvert \xi  - x \rvert \leq \frac{1}{\sqrt{i+1}} \right] \right)$
		\State $d = \min \left( \hat{f}^{-1} , d_0 (i+1)^{a} \right) $
		\State $\textrm{state}.d \gets d$
		\State $\textrm{state}.i \gets i+1$
		\State $\textrm{state}.\xi \gets \xi$
		\State $\textrm{state}.\hat{f} \gets \hat{f}$
		\State\Return \textrm{state} 
		\EndFunction
	\end{algorithmic}
\end{algorithm}

\begin{algorithm}
	\caption{SCAPA algorithm}
	\label{alg:OCAPA}
	\begin{algorithmic}[1]
		\Inputs{Penalty parameter $\lambda$ \\
			A minimum segment length $l \geq 2$ \\
			A maximum segment length $m > l$\\
			Burn in period $n_0 > l$ \\ 
			A cost function $\mathcal{C}(\cdot)$, such as twice the minimised negative log-likelihood for a segment of data $x_{(k+1):t}$, i.e.\ 
			\begin{align*}
			\mathcal{C}(x_{(k+1):t}) = (t-k)\left[ \log \left( \frac{1}{t-k} \sum_{i=k+1}^{t}(x_i - \bar{x}_{(k+1):t})^2 \right) + 1 \right]. 
			\end{align*}
		}
		\Initialize{ Allocate $x_{1:n_0}$ to the typical distribution \Comment{Burn in period} \\
			$\mathrm{state_{lq}} \gets \Call{InitialQuantile}{x_{1:n_0},0.25}$ \\ 
			$\mathrm{state_{med}} \gets \Call{InitialQuantile}{x_{1:n_0},0.5}$ \\
			$\mathrm{state_{uq}} \gets \Call{InitialQuantile}{x_{1:n_0},0.75}$ \\
			$\hat{\mu} \gets \mathrm{state_{med}.\xi}$ \\
			$\hat{\sigma} \gets \frac{1}{2\Phi^{-1}(0.75)} \left( \mathrm{state_{uq}.\xi} - \mathrm{state_{lq}.\xi} \right)$ \\
			$C(t) \gets \frac{ 1 }{ \hat{\sigma}^2 } \sum_{i=1}^{t} (x_i - \hat{\mu} )^2 \hspace{10pt} \forall t = 1,2, \hdots ,n_0$  \\
			$Anom(t) = NULL \hspace{10pt} \forall t = 1,2, \hdots ,n_0$  \\
			$t \gets n_0 + 1$
		}
		\While{TRUE}\Comment{Until no longer observe any new data}
		
		\State Compute updated estimate of median $\hat{\mu}$ and $\hat{\sigma}$ using $x_t$
		\State $ \mathrm{state_{lq}} \gets \Call{UpdateQuantile}{\mathrm{state_{lq}}, x_t , 0.25 }$
		\State $ \mathrm{state_{med}} \gets \Call{UpdateQuantile}{\mathrm{state_{med}}, x_t , 0.5 }$
		\State $ \mathrm{state_{uq}} \gets \Call{UpdateQuantile}{\mathrm{state_{uq}}, x_t , 0.75 }$
		\State $\hat{\mu} \gets \mathrm{state_{med}.\xi}$ 
		\State $\hat{\sigma} \gets \frac{1}{2\Phi^{-1}(0.75)} \left( \mathrm{state_{uq}.\xi} - \mathrm{state_{lq}.\xi} \right)$ 
		\State $x_{t} \leftarrow  \frac{x_{t} - \hat{\mu}}{\hat{\sigma}}$ \Comment{Centralise new data point}
		\State $C_1(t) \gets C(t-1) + x_t^2$ 
		\State $C_2(t) \gets C(t-1) + 1 + \log \left( \gamma + x_t^2 \right) + \beta_O(\lambda)$
		\State $C_3(t) \gets \min_{ t - m \leq k \leq t - l} \left[ C(k) + \mathcal{C}(x_{(k+1):t}) + \beta_C(t-k,\lambda) \right]$
		\State $s \gets \argmin_{ t - m \leq k \leq t - l} \left[ C(k) + \mathcal{C}(x_{(k+1):t}) + \beta_C(t-k,\lambda) \right]$ 
		\State $C(t) \gets \min [ C_1(t) , C_2(t) , C_3(t) ]$
		\Switch{ $\argmin [C_1(t),C_2(t),C_3(t)] $ }
		\Case{$1$:}
		\State $Anom(t) \gets Anom(t-1)$ \Comment{$x_t$ is from the typical distribution}  
		\EndCase
		\Case{$2$:}
		\State $Anom(t) \gets [Anom(t-1),(t)]$ \Comment{$x_{t}$ is a point anomaly}
		\EndCase
		\Case{$3$:}
		\State $Anom(t) \gets [Anom(s),(s+1,t)]$ \Comment{$x_{(s+1):t}$ is a collective anomaly}   
		\EndCase
		\EndSwitch
		\State $t \gets t + 1$ 
		\EndWhile
	\end{algorithmic}
\end{algorithm}

\newpage

\subsection{Proofs}

\subsubsection{Proof of Proposition \ref{Thm:ARL}}

Without loss of generality assume that $\mu_0 =0$ and $\sigma_0 =1$. Let $T$ be the time at which the first anomaly is declared. By definition of the ARL and of the test statistic used by SCAPA 
\begin{align*}
ARL = \sum_{n=0}^{\infty} \mathbb{P} \left( T > n \right)
= \sum_{n=0}^{\infty} \mathbb{P} \left( \max_{1 \leq a \leq b \leq n, b - a < m} (b-a+1)\left(\bar{x}_{a:b}\right)^2  < \lambda \right),
\end{align*}
where $\bar{x}_{a:b}$ denotes the arithmetic mean of $x_a,...,x_b$. The ARL is therefore bounded above by
\begin{align*}
&\sum_{n=0}^{\infty} \mathbb{P} \left( (a-a+1)\bar{x}_{a:a}^2  < \lambda , \;\;\; 1 \leq a \leq n \right) = \sum_{n=0}^{\infty} \prod_{a=1}^{n} \mathbb{P}\left( \left(x_a\right)^2  < \lambda  \right) = \sum_{n=0}^{\infty} \mathbb{P}\left( \chi^2_1  < \lambda  \right)^n \\
& = \mathbb{P}\left( \chi^2_1  > \lambda  \right)^{-1} = \frac{1}{2\mathbb{P}\left( N(0,1) > \sqrt{\lambda}  \right) } \leq \sqrt{\frac{\pi}{2}}\frac{\lambda + 1}{\sqrt{\lambda}} e^{\frac{\lambda}{2}},
\end{align*}
where the inequality follows from tail bounds on the normal distribution. Furthermore, for any $\epsilon>0$ we can bound the ARL from below by 
\begin{align*}
\sum_{n=0}^{\lfloor e^{\frac{\lambda(1-\epsilon)}{2(1+\epsilon)}} \rfloor - 1} \mathbb{P} \left( T > n \right)
\geq \sum_{n=0}^{\lfloor e^{\frac{\lambda(1-\epsilon)}{2(1+\epsilon)}} \rfloor - 1} \mathbb{P} \left( \max_{1 \leq a \leq b \leq n} (b-a+1)\left(\bar{x}_{a:b}\right)^2  < \lambda \right).
\end{align*}
Lemma 6 from \cite{2019MVCAPA} shows that there exists a universal constant $\tilde{A}$ such that the above exceeds
\small
\begin{align*}
&\sum_{n=0}^{\lfloor e^{\frac{\lambda(1-\epsilon)}{2(1+\epsilon)}} \rfloor - 1}\left[ 1 - \tilde{A}(n+1)\log(n+1) \frac{1}{\log(1+\epsilon)} e^{-\frac{\lambda}{2(1+\epsilon)}}\right] 
= \lfloor e^{\frac{\lambda(1-\epsilon)}{2(1+\epsilon)}} \rfloor - \frac{\tilde{A} e^{-\frac{\lambda}{2(1+\epsilon)}}}{\log(1+\epsilon)} \sum_{n=1}^{\lfloor e^{\frac{\lambda(1-\epsilon)}{2(1+\epsilon)}} \rfloor} n\log(n) \\
& \geq e^{\frac{\lambda(1-\epsilon)}{2(1+\epsilon)}} - 1 - \frac{\tilde{A} e^{-\frac{\lambda}{2(1+\epsilon)}}}{\log(1+\epsilon)}  \left(\lfloor e^{\frac{\lambda(1-\epsilon)}{2(1+\epsilon)}} \rfloor\right)^2 \log  \left(\lfloor e^{\frac{\lambda(1-\epsilon)}{2(1+\epsilon)}} \rfloor\right) \\
&\geq e^{\frac{\lambda(1-\epsilon)}{2(1+\epsilon)}} - 1 -  \frac{\tilde{A} }{\log(1+\epsilon)}   e^{\frac{\lambda(1-2\epsilon)}{2(1+\epsilon)}}  \log  \left( e^{\frac{\lambda(1-\epsilon)}{2(1+\epsilon)}} \right),  
\end{align*}
which exceeds $e^{\frac{\lambda(1-\epsilon)^2}{2(1+\epsilon)}}$ for sufficiently high values of $\lambda$. This finishes the proof.

\subsubsection{Proof of Proposition \ref{Thm:ADD_max}}
For a maximum segment length $m$, define the stopping time $T_m$, to be the first time at which a collective anomaly was detected. Define $T_{\infty}$ to be the stopping time of SCAPA without maximum segment length. Note that $ADD_m = \mathbb{E}\left( T_m \right)$ and $ADD_\infty = \mathbb{E}\left( T_\infty \right)$. Clearly, $ADD_m$ decreases in $m$ and $ADD_m \geq ADD_\infty$ for all $m \geq 1$. It is therefore sufficient to show that $ADD_m < ADD_\infty + o(1)$ for $m = \lceil\frac{\lambda}{\mu^2}(1+\epsilon) \rceil$. We have that
\begin{align*}
ADD_m &= \sum_{i=1}^{\infty} \mathbb{P} \left( T_m \geq i\right) = \sum_{i=1}^{m} \mathbb{P} \left( T_m \geq i\right) + \sum_{j=1}^{\infty}\left[\sum_{i=mj+1}^{m(j+1)} \mathbb{P} \left( T_m \geq i\right) \right]  \\
&= \sum_{i=1}^{m} \mathbb{P} \left( T_\infty \geq i\right) + \sum_{j=1}^{\infty}\left[\sum_{i=mj+1}^{m(j+1)} \mathbb{P} \left( T_m \geq i\right) \right] \leq ADD_\infty + \sum_{j=1}^{\infty}m\mathbb{P} \left( T_m \geq mj\right) \\
&\leq ADD_\infty + \sum_{j=1}^{\infty}m\mathbb{P} \left( T_m \geq m\right)^{j} = ADD_{\infty} + m \mathbb{P} \left( T_m \geq m\right) \frac{1}{1-\mathbb{P} \left( T_m \geq m\right)}.
\end{align*}
Note that the above is strictly increasing in $\mathbb{P} \left( T_m \geq m\right)$. Writing $x_t=\mu+\eta_t$, where $\eta_t \sim N(0,1)$, we can bound this probability by: 
\begin{align*}
\mathbb{P} \left( T_m \geq m\right) &< \mathbb{P} \left( \sqrt{m}|\mu+\bar{\eta}_{1:m}|  \leq \sqrt{\lambda} \right) \leq \mathbb{P} \left( \sqrt{m}\bar{x\eta}_{1:m}  \leq \sqrt{\lambda} - \sqrt{m}|\mu| \right) \\
&\leq \mathbb{P} \left(N(0,1) 
\leq -\sqrt{\lambda}(\sqrt{1+\epsilon} - 1) \right) 
\leq \exp \left(- \frac{1}{2}(\sqrt{1+\epsilon} - 1)^2 \lambda \right).
\end{align*}
Here, the inequality follows form standard tail bounds on the normal distribution. Consequently, 
\begin{equation*}
m \mathbb{P} \left( T_m \geq m\right) \frac{1}{1-\mathbb{P} \left( T_m \geq m\right)} \leq 
\left(1+(1+\epsilon)\frac{\lambda}{\mu^2}\right) \mathbb{P} \left( T_m \geq m\right) \frac{1}{1-\mathbb{P} \left( T_m \geq m\right)}
= o(1),
\end{equation*}
which finishes the proof.

\subsubsection{Proof of Proposition \ref{Thm:ADD_min}}

The proof 
follows that of Proposition \ref{Thm:ARL}. We have that 
\begin{align*}
ADD_m &= \sum_{n=0}^{\infty} \mathbb{P} \left( T_m > n \right) \leq \sum_{n=0}^{\infty} \mathbb{P} \left( (a-a+1)\bar{x}_{a:a}^2  < \lambda , \;\;\; 1 \leq a \leq n \right)   \\
&\leq \sum_{n=0}^{\infty} \mathbb{P} \left( \bar{\epsilon}_{a}^2  < \lambda , \;\;\; 1 \leq a \leq n \right)  = \sum_{n=0}^{\infty} \mathbb{P} \left( \chi^2_1  < \lambda \right)^n. \\
& = \mathbb{P}\left( \chi^2_1  > \lambda  \right)^{-1} = \frac{1}{2\mathbb{P}\left( N(0,1) > \sqrt{\lambda}  \right) } \leq \sqrt{\frac{\pi}{2}}\frac{\lambda + 1}{\sqrt{\lambda}} e^{\frac{\lambda}{2}}.
\end{align*}
Further, for any $\delta>0$ we can write $x_t=\mu+\eta_t$, where $\eta_t \sim N(0,1)$/ Consequently, we can bound the ADD from below by 
\begin{align*}
&\sum_{n=0}^{\lfloor e^{\frac{\lambda(1-\delta)}{2(1+\delta)}} \rfloor - 1} \mathbb{P} \left( \max_{1 \leq a \leq b \leq n, b-a+1 \leq m} \sqrt{(b-a+1)}\left|\mu+\bar{\eta}_{a:b}\right|  < \sqrt{\lambda} \right) \\
& \geq \sum_{n=0}^{\lfloor e^{\frac{\lambda(1-\delta)}{2(1+\delta)}} \rfloor - 1} \mathbb{P} \left( \max_{1 \leq a \leq b \leq n, b-a+1 \leq m} \sqrt{(b-a+1)}|\mu|+ \sqrt{(b-a+1)}\left|\bar{\eta}_{a:b}\right|  < \sqrt{\lambda} \right) \\
& \geq \sum_{n=0}^{\lfloor e^{\frac{\lambda(1-\delta)}{2(1+\delta)}} \rfloor - 1} \mathbb{P} \left( \max_{1 \leq a \leq b \leq n, b-a+1 \leq m} \sqrt{(b-a+1)}\left|\bar{\eta}_{a:b}\right|  < \sqrt{\lambda} - |\mu|\sqrt{m} \right) \\
& \geq \sum_{n=0}^{\lfloor e^{\frac{\lambda(1-\delta)}{2(1+\delta)}} \rfloor - 1} \mathbb{P} \left( \max_{1 \leq a \leq b \leq n} \sqrt{(b-a+1)}\left|\bar{\eta}_{a:b}\right|  < \sqrt{\lambda} - |\mu|\sqrt{m} \right).
\end{align*}
Replicating the arguments in the proof of Proposition \ref{Thm:ARL}, this can be shown to exceed
\begin{align*}
e^{\frac{\tilde{\lambda}(1-\delta)}{2(1+\delta)}} - 1 -  \frac{\tilde{A} }{\log(1+\delta)}   e^{\frac{\tilde{\lambda}(1-2\delta)}{2(1+\delta)}}  \log  \left( e^{\frac{\tilde{\lambda}(1-\delta)}{2(1+\delta)}} \right),  
\end{align*}
where $\tilde{\lambda} := \left(\sqrt{\lambda} - |\mu|\sqrt{m}\right)^2 \rightarrow \lambda$. This finishes the proof. 